\documentclass[aps,prb,twocolumn,superscriptaddress,showpacs,eqsecnum,floatfix]{revtex4-1}
\usepackage{amssymb,amsmath}
\usepackage{graphicx}
\usepackage{bm}
\usepackage{subfigure,graphicx}
\begin{document}

\title{Inhomogeneous superconducting phases in the frustrated Kondo-Heisenberg chain}
\author{Ariel Dobry}
\affiliation{
Facultad de Ciencias Exactas Ingenier{\'\i}a y Agrimensura, Universidad Nacional de Rosario, and Instituto de
F\'{\i}sica Rosario, Bv. 27 de Febrero 210 bis, 2000 Rosario, Santa F\'e,
Argentina.}
\author{Akbar Jaefari}
\affiliation{Department of Physics and Astronomy, University of Oklahoma, Norman, Oklahoma 07319, USA}
\affiliation{
Department of Physics, University of Illinois at Urbana-Champaign, University of Illinois at Urbana-Champaign,
1110 West Green Street, Urbana, Illinois 61801-3080, USA }
\author{Eduardo Fradkin}
\affiliation{
Department of Physics and Institute for Condensed Matter Theory, University of Illinois at Urbana-Champaign, 1110 West Green Street, Urbana, Illinois 61801-3080, USA}
\date{\today}
%%%%%%%%%%%%%%%%%%%%%%%%%%%%%%%%%%%%%%%%%%%%%%%%%%%%%%%%%

\begin{abstract}
We use bosonization and renormalization group methods to determine the ground state phase diagram of a one-dimensional frustrated Kondo-Heisenberg system 
consisting of a one-dimensional spin-1/2 Luttinger liquid coupled by a Kondo exchange interaction $J_K$ to a frustrated quantum antiferromagnetic Heisenberg chain, 
with a nearest-neighbor exchange coupling $J_1$ and a next-nearest-neighbor (frustrating) exchange interaction $J_2$. 
We analyze the interplay of quantum frustration in the antiferromagnetic chain with the Kondo exchange coupling $J_K$ with the Luttinger liquid. 
We discuss the structure of the phase diagram of this system as a function of the ratios $J_K/J_1$, $J_2/J_1$ and of the parameters of the Luttinger liquid. 
In particular we discuss in detail the regimes in which a pair-density-wave state may be realized and its relation with the spin correlations
in the frustrated antiferromagnetic chain.
\end{abstract}
\maketitle

\section{Introduction}

Kondo-Heisenberg chains are simple model systems in which a (one-dimensional) Luttinger liquid is coupled to a one-dimensional antiferromagnetic system 
by a Kondo exchange interaction. Systems of this type display a variety of non-trivial ground states with a dazzling array of 
unconventional behaviors.\cite{zachar-1996,white-1996,sikkema-1997,andrei-2000,zachar-2001,zachar-2001b,berg-2010} 
In particular,  they have phases that exhibit a spin gap and have superconducting correlations with the peculiar aspect  that their only viable order parameters 
are made of observables from the spin chain and of the Luttinger liquid which separately have only short range correlations. Kondo-Heisenberg systems constitute an ideal 
testing ground in which the physics of strongly correlated electronic systems can be studied using controlled analytic approximations and powerful numerical methods. 

One of the phases that is well established to exist in the Kondo-Heisenberg chain is  a state with a spin gap and long-range (power-law) superconducting 
correlations with finite momentum (which is commensurate with the lattice of Kondo-Heisenberg spins). This state was identified in Ref. [\onlinecite{berg-2010}] 
as a pair-density-wave superconducting (PDW) state,  i.e. a Larkin-Ovchinnikov state without an external Zeeman (magnetic) field. 
In this paper we will consider the effects of next-nearest-neighbor antiferromagnetic exchange interactions in the spin chain. In the presence of such 
interactions the spin chain is frustrated and can be regarded as a zig-zag spin ladder. 
The purpose of this paper is to determine the phase diagram of this system as a function of the ratio of the Kondo exchange interaction $J_K$ to the 
nearest-neighbor exchange interaction $J_1$ in the spin chain and of the ratio between the next-nearest-neighbor exchange interaction of the spin chain $J_2/J_1$, 
as well as the interaction coupling constants of the Luttinger liquid. 

One motivation for considering the role of frustration is to examine the mechanisms that determine the ordering wave-vector $Q$ of a pair-density-wave state. 
In a one-dimensional fermionic system with at least two bands (or species of fermions) with attractive interactions it is possible to have a Larkin-Ovchinnikov state 
with any wave vector provided that the Fermi points of the two species are different. This is possible since in 1D there is always nesting of the Fermi points. 
However,  if the strongly correlated system has only repulsive interactions this problem is more subtle. For example, Berg and coworkers \cite{berg-2010} 
found that in Kondo-Heisenberg chain the ordering wave-vector $Q$ of the PDW is the same as the wave-vector of the antiferromagnetic order, which in turn is 
determined by the spacing of the Kondo spins. This is true even though in such a spin-gap phase the magnetic order is short ranged. 
A similar result was obtained in Ref. [\onlinecite{jaefari-2012}] in the context of doped antiferromagnetic two-leg spin ladders, with the only difference that the ordering 
wave vector is determined by spontaneous-symmetry-breaking of translation invariance rather than the explicit breaking of translation invariance of the 
Kondo-Heisenberg chain. Thus, in both cases the PDW order is commensurate. 

It is apparent that this commensurate state is the result of the magnetic origin of the  mechanism of the PDW order,  provided the SU(2) magnetic symmetry is exact. 
This is true even in the case of the frustrated chain which, for $J_2$ larger than a critical value, at the classical level the frustrated Heisenberg chain  has a ground state with incommensurate spiral 
magnetic order. On the other hand, at the quantum level, in the case of a chain of spin-1/2 degrees of freedom the ground state is dimerized and has short-ranged spin order.\cite{white-1996} 
The short range magnetic order becomes incommensurate order for $J_2>J_1/2$, the Majumdar-Ghosh point\cite{majumdar-1969}, where the dimerization is strongest.

This behavior naturally poses the question of whether in a strictly one-dimensional system an incommensurate phase is possible at all. A ground stye with incommensurate magnetic order requires the existence of an additional gapless collective mode. In a one-dimensional system this is possible only if the effective low energy theory 
has an exactly marginal operator. However, the effective field  theories of models with an exact SU(2) symmetry (and their generalizations) 
are non-linear sigma models with compact target space manifolds. these target manifolds do not have flat directions and hence their ground states can only have massive (gapped) excitations.\cite{friedan-1985} 
The only alternative to this gapped ground state is that the generalized non-linear sigma models may have a topological term. In this case the RG flows drives the system to a non-trivial finite 
conformally invariant fixed point. 
The only available fixed points with SU(2) symmetry are Wess-Zumino-Witten (WZW) models with current algebras SU(2)$_k$ (where $k$ is an integer) 
(or generalizations of these fixed points). However, fixed points of this class also do not have any exactly marginal operators. These arguments suggest that it is not possible to have a state with incommensurate magnetic order with exact SU(2) spin invariance in one spatial dimension.
The only way to circumvent this problem is either to break the magnetic symmetry explicitly, from SU(2) down to a U(1) subgroup, as in the presence of 
magnetic anisotropy, which allows for magnetic spiral phases,\cite{sato-2011} or to consider a quasi-one-dimensional system in which the symmetry-breaking is 
spontaneous in the higher-dimensional system (or by an explicit symmetry-breaking by an uniform magnetic field)\cite{starykh-2007,schnyder-2008,starykh-2010} 
Nevertheless, although magnetic frustration in an SU(2) invariant system does not lead to an incommensurate PDW phase in the 1D case we are discussing here, 
we will see that it does lead to non-trivial and interesting phases that will be discussed in some detail in this paper.

In this paper we investigate the nature of the ground state of a frustrated Kondo-Heisenberg model and the resulting phase diagram (see Fig. \ref{fig:diagram}).
 Our results are based on a combination of bosonization methods (both abelian and non-abelian) with perturbative renormalization group arguments. Our strategy is to investigate the behavior of the system in limiting regimes of the parameter space. Thus we will first consider the extension of the results of Ref. [\onlinecite{berg-2010}] account for the effects of weak next-nearest neighbor interactions $J_2$ in the spin chain. Next we consider the opposite regime in which the nearest-neighbor interactions $J_1$ of the spin chain are weak. This regime turn out to be quite rich. Finally we consider the regime in which $J_1$ is weak compared with the Kondo exchange coupling. In this regime we find a finite non-trivial fixed point of the Toulouse type, investigated earlier on for a somewhat different system by Azaria and Lecheminant.\cite{azaria-2000} By considering the leading perturbations around the Toulouse fixed point we show that it controls a stable phase with the character of a fractionalized spin liquid. At some finite value of $J_1$ this phase becomes unstable and has a phase transition to a pair-density-wave phase.

This paper if organized as follows. In Section \ref{sec:model} we introduce the frustrated Kondo-Heisenberg model 
and discuss how the order parameters of the different phases of interest are realized. 
In Section \ref{sec:J2llJ1} we consider the weak frustration regime, $J_2 \ll J_1$, which is treated using bosonization methods (both abelian and non-abelian) 
and renormalization group calculations. Here we discuss the interplay between the PDW phase of the unfrustrated system and the dimerized phase. 
In Section \ref{sec:J2ggJ1} we discuss the opposite regime, $J_2 \gg J_1$ which can also be treated by bosonization and renormalization group methods. 
In this regime the phase diagram turns out to be quite rich and the construction of the order parameters is non-trivial. Here we find a stable phase with a gapless fractionalized fluid in the spin sector, a fractionalized spin liquid phase,  and discuss the quantum phase transition from this phase to the PDW phase of the weakly frustrated regime.
We close this paper  by discussing the resulting phase diagram of this system, shown in Fig. \ref{fig:diagram},  in Section \ref{sec:conclusions} as well as several open questions. The Toulouse point solution of the strong coupling regime of this system is summarized in Appendix \ref{sec:toulouse}.

\section{The Model}
\label{sec:model}

The frustrated Kondo-Heisenberg chains a model of a one-dimensional interacting system of spin-1/2 fermions coupled by a Kondo exchange coupling to a 
one-dimensional array of localized spin-1/2 degrees of freedom whose Hamiltonian is a Heisenberg antiferromagnet with both nearest- and 
next-nearest neighbor exchange interactions. When the latter interactions are larger than a critical value the resulting magnetic chain is frustrated and 
can equivalently be depicted as a zig-zag ladder.
%We add a nearest next neighbor to the Heisenberg chain of the problem treated in Ref. \onlinecite{berg-2010} 

The model is described by the  Hamiltonian of Eq.\eqref{Htotal} which is a sum of three terms: a)  $H_F$ that represents the interacting system of fermions, 
Eq.\eqref{H1DEG}, b) $H_{Heis}$ for the frustrated Heisenberg, Eq.\eqref{hheis} , and c) $H_K$ that describes the Kondo exchange interaction between the 
spin degrees of freedom in these two subsystems, Eq.\eqref{hhfruskondo},
\begin{eqnarray}
H&=&H_F+H_{Heis}+H_K
\label{Htotal}\\
H_{F}&=&-t \sum_{i,\sigma}(c^{\dagger}_{i\sigma} c_{i+1 \sigma}+
c^{\dagger}_{i+1\sigma} c_{i \sigma})+H_{\rm int}
%+U \sum_{i,\sigma}n_{i\uparrow}n_{i\downarrow} \\
\label{H1DEG}\\
H_{Heis}&=&J_1 \sum_{i}\mathbf{S}_{i}\cdot \mathbf{S}_{i+1}+J_2 \sum_{i,\sigma}\mathbf{S}_{i}\cdot \mathbf{S}_{i+2}
\label{hheis}\\
H_{K}&=&J_K \sum_{i}\mathbf{S}_{i} \cdot \mathbf{s}_{i}
 \label{hhfruskondo}
\end{eqnarray}
In Eq.\eqref{H1DEG} $H_{\rm int}$ represent local interactions in the system of 1D fermions, {\it e.g.} a Hubbard interaction $U$, a nearest-neighbor (Coulomb) 
repulsion $V$, and a nearest-neighbor (Heisenberg) exchange interaction $J$. Here we will assume that the 1D system of fermions is gapless and hence described 
by a Luttinger liquid with charge and spin excitations. Thus, the 1D fermionic system has a U(1) $\times$ SU(2) global symmetry (accounting for the charge and spin sectors). 
The Heisenberg spin chain has an SU(2) global symmetry. 
The Kondo coupling between the two systems, represented by $H_K$ in Eq.\eqref{hhfruskondo} with coupling constant $J_K$,  reduces the symmetry to a 
U(1) $\times$ SU(2) global symmetry. In principle the lattice spacing of the magnetic chain is different than that of the 1D system of mobile fermions. 
For simplicity here we will take the lattice spacings  of the two subsystems to be the same. In this case there is no explicit breaking of translation invariance.

In Eq.\eqref{H1DEG} we introduced a set of fermion creation and annihilation operators, $c^\dagger_{i,\sigma}$ and $c_{i,\sigma}$, 
at each site $i$ for electrons with spin $\sigma=\uparrow,\downarrow$. The three components of the spin operator of the itinerant electrons at the $i$th site are given by
$\mathbf{s}^a_i=\frac12 \sum_{\alpha,\beta}c^{\dagger}_{i,\alpha}\mathbf{\sigma}^a_{\alpha,\beta}c_{i,\beta}$, where $a=1,2,3$. 
Here ${\sigma}^a_{\alpha,\beta}$ is the 
$(\alpha,\beta)$ element of the
$a$th Pauli matrix. Similarly, the operators $\mathbf{S}_i$ represent the spins of the frustrated Heisenberg spin chain with coupling constants $J_1$ 
(for the nearest-neighbor exchange coupling) and $J_2$ for the next-nearest-neighbor exchange coupling. 

As we noted before the frustrated chain can also be regarded as a zig-zag two-leg spin ladder with $J_2$ representing the interactions on the rungs. 
In this picture, the frustrated chain is a 1D version of an asymmetric triangular lattice. This picture is appropriate for quasi-one-dimensional systems such as 
Cs$_2$CuCl$_4$.\cite{coldea-2001} Both coupling constants are taken to be antiferromagnetic. Hence the spin chain is frustrated. 
On the other hand, in what follows we will take the Kondo coupling to be the same for all sites which is appropriate for the frustrated chain.

Several limits of this model have been considered before. In the case in which the Kondo coupling is zero, $J_K=0$, it is known \cite{haldane-1982} 
that the frustrated chain has a quantum phase transition
at a critical value $J_{2c}$ of $J_2$. For $J_2<J_{2c}$, frustration has essentially not effect on the low energy properties. In this regime, next-nearest-neighbor 
interaction is marginally irrelevant and the system behaves effectively as the spin-1/2 Heisenberg model chain. 
Thus, in this regime the decoupled spin chain is gapless, and hence critical, and exhibits power law correlations. 
When $J_2$ exceeds a critical value of the next-nearest-neighbor coupling $J_{2c}$, the interaction becomes marginally relevant. 
The resulting ground state is a spin singlet and there is a finite energy gap in the spectrum.\cite{white-1996} 
In this regime the translational order is  spontaneously broken by the appearance of a magnetic dimerization of the chain. The order parameter for the dimerized phase is  
\begin{equation}
\epsilon\equiv(-1)^i \left\langle \mathbf{S}_i\cdot\mathbf{S}_{i+1}-\mathbf{S}_{i+1}\cdot\mathbf{S}_{i+2}\right\rangle
\label{eq:dimerization}
\end{equation}
 which acquires a nonzero, position-independent value in this phase. The value $\frac{J_{2c}}{J_1}=0.241$ is known from numerical studies. 
 At $\frac{J_{2c}}{J_1}=0.5$, known as a Majundar-Gosh point, the ground state could be analytically obtained. 
 It corresponds to the formation of spin singlets between each site and one of its neighbors.  In this limit the dimerization (and the frustration) is largest. 
 Inside the dimerized phase there is short-range incommensurate spiral-spin order. Since there is a finite spin gap, not only electron tunneling between the two chains 
 is suppressed, but also the exchange Kondo coupling becomes irrelevant up to a finite value of the order of the spin gap. 
 Hence in this regime the gapless 1D electron system and the frustrated spin chain are effectively decoupled at low energies. 

On the other hand, for $J_2=0$ and $J_K\neq 0$, the system is the Kondo-Heisenberg chain. For small $J_K$ the Kondo coupling 
is a marginally relevant perturbation and flows under the RG to a strong coupling fixed point with a finite spin gap 
and short-range commensurate magnetic correlations.\cite{sikkema-1997} 
In this system only composite order parameters have power-law correlations.\cite{zachar-1996,zachar-2001,zachar-2001b} 
It has been realized recently\cite{berg-2010} that in the spin gap phase (usually referred to as the ``Kondo singlet'' regime in the heavy fermion literature) 
the strongest correlations (i.e. with the smallest exponent) describe a pair-density-wave state, a superconducting order with finite wave-vector. 
In the case in which the two lattice spacings are the same the ordering wave vector is $Q=\pi$.

Finally, the limit in which $J_1=0$ but with $J_2$ and $J_K$ finite has also been studied in some detail. This limit is quite rich. It has an unstable fixed point at $J_K=0$. 
For finite $J_K$  this system flows to a non-trivial finite infrared stable fixed point which in some ways resembles the physics of the 
multi-channel Kondo problem.\cite{azaria-2000,azaria-2000b,lecheminant-2002} This fixed point is equivalent to two decoupled chirally stabilized systems.\cite{andrei-1998} 
In the phase governed by this non-trivial fixed point the only allowed order parameters (i.e. operators with power-law correlations) are also composite.

In the subsequent sections we will examine the phase diagram of the full system by expanding about these two limits, $J_2\ll J_1$ and $J_2\gg J_1$, with $J_K$ finite 
(and with the parameters of the 1D gapless electronic system fixed).
%introduction of the PDW phase in the case J_2=0

\section{$J_2<<J_1$}
\label{sec:J2llJ1}

In this section we will look at the effects of the next-nearest-neighbor exchange coupling $J_2$ on the physics Kondo-Heisenberg chain. 
In the regime with small $J_K$ and small $J_2$ one can look at the effective field theory in a naive continuum limit and analyze the role of various operators. 
Bosonization (abelian and non-abelian) are very useful tools to understand the physics of this regime.\cite{zachar-1996,zachar-2001,zachar-2001b}
The low energy excitations of the electronic chain can be taken into account by linearizing the free fermion band around the Fermi level. 
In this regime the fermionic operators can be written in terms of the continuum right and left fields 
$\psi^1_{\sigma R}(x)$ and $\psi^1_{\sigma L}(x)$ (here the label $1$ denotes the 1D system of mobile electrons) as :
\begin{eqnarray}
\frac{c_{n \sigma}}{\sqrt{a}}&\sim& e^{ik_F n} \psi_{\sigma R,1}(x) +e^{-ik_F n}\psi_{\sigma L,1}(x)  
\label{fermion}
\end{eqnarray}
where $a$ is the lattice spacing and $k_F$ is the Fermi momentum of the chain.
The local spin operator of the electrons $\mathbf{s}_n=\frac12 c^{\dagger}_{n\alpha}\mathbf{\sigma}_{\alpha \beta} c_{n \beta}$,  
can also be decomposed into a slowly varying piece (with Fourier components near zero wave vector) and a rapidly varying piece 
(with Fourier components near $2k_F$) which represents the spin-density wave order parameter. More explicitly 
\begin{eqnarray}
\mathbf{s}_n &\sim& a \left[ \mathbf{J}_{L,1}(x)+\mathbf{J}_{R,1}(x)+ e^{-i2k_Fn} \mathbf{N}_{1}(x)+ \textrm{h.c.}\right]
\label{spinel}
\end{eqnarray}
where
\begin{align}
\mathbf{J}_{R,1}&=\frac12 \psi^{\dagger}_{\alpha R,1} \boldsymbol{\sigma}_{\alpha,\beta}\psi_{\beta R,1}\nonumber\\
\mathbf{J}_{L,1}&=\frac12 \psi^{\dagger}_{\alpha L,1} \boldsymbol{\sigma}_{\alpha \beta}\psi_{\beta L,1}
\label{spin-currents-1DEG}
\end{align}
are the chiral spin currents of the right and left moving electrons, and
\begin{eqnarray}
\mathbf{N}_1=\frac12 \psi^{\dagger}_{\alpha R,1} \boldsymbol{\sigma}_{\alpha\beta}\psi_{\beta L,1}
\label{defNe}
\end{eqnarray}
For general $k_F$ the
order parameter operator $\mathbf{N}_1$ of the 1D electronic system is a  three-component complex vector and, for general filling of the 1D electronic system, 
the spin-density wave has an ordering wave-vector $Q=2k_F$. In the special case in which the 1D electronic chain is half-filled, where there is a Mott charge gap, the Kondo-Heisenberg chain is a Kondo insulator. In this case the ordering wave vector of the 1D electronic system is $Q=\pi$ and $\mathbf{N}_1$ is real 
(self-adjoint) and is a N\'eel order parameter. We will not discuss this interesting case here and we will focus on the case in which the electronic system is metallic.

%In this work we assume  $J_K<<J_1, J_2, t$. We start by studying the weak frustrated case where $J_2<<J_1$  
On the other hand, the degrees of freedom of the Heisenberg spin chain can also be decomposed into slowly and rapidly varying components.\cite{affleck-1986}
In this case the spin operators of the Heisenberg chain could also be represented by fermionic fields starting from a half filled Hubbard model (with $k_F=\pi$) 
and a gapped (and hence  frozen) charge degrees of freedom (due to the Mott gap). The decomposition of the local spin operators of the Heisenberg chains is: 
\begin{eqnarray}
\mathbf{S}_n &\sim& a \left[ \mathbf{J}_{L,2}(x)+\mathbf{J}_{R,2}(x)+ (-1)^n \mathbf{N}_2(x)\right]
\label{spin2}
\end{eqnarray}
where the label $2$ now denotes the spin chain.
The expressions for the chiral spin currents of the Heisenberg chain $\mathbf{J}_{L,2}$ and $\mathbf{J}_{R,2}$  
are the same as the previous ones by changing 1 for 2. The (real) N\'eel order parameter (the staggered magnetization) is:
\begin{eqnarray}
\mathbf{N_2}=\frac12 \left[\psi^{ \dagger}_{\alpha R,2} \bm{\sigma}_{\alpha,\beta}\psi_{\beta L,2}+ \psi^{ \dagger}_{\alpha L,2} \bm{\sigma}_{\alpha,\beta}\psi_{\beta R,2}\right]
\end{eqnarray}

The effective Hamiltonian for the low energy regime can be determined from the Hamiltonian of Eqs.\eqref{H1DEG}, 
\eqref{hheis}, and \eqref{hhfruskondo}. As it is common in 1D problems, here too in the low energy regime  there is a separation 
between charge and spin degrees of freedom and the effective Hamiltonian is a sum of two terms, one for the charge sector 
and one for the spin sector. (for a general discussion on bosonization see, e.g., Ref. [\onlinecite{fradkin-2013}].) 
The charge sector is described  by a conventional (compactified) boson $\phi_c$ whose Hamiltonian density, $\mathcal{H}_c$, 
is parametrized by a charge Luttinger parameter  $K_c$ and a charge velocity $v_c$, both of which depend in a non-universal way 
on the microscopic parameters of Eqs.\eqref{H1DEG}, \eqref{hheis}, and \eqref{hhfruskondo}
\begin{equation}
\mathcal{H}_c= \frac{v_c}{2K_c} \Pi_c^2+\frac{1}{2}K_c v_c (\partial_x \phi_c)^2
\label{Hcharge}
\end{equation}
where $\Pi_c$ is momentum canonically conjugate to the field $\phi_c$. The (normal-ordered) charge density $j_0$ and charge current are related to the field $\phi_c$ 
by the usual  bosonization formula.

 The Hamiltonian density for the spin sector, $\mathcal{H}_s$, can be written in terms of the right and left moving spin currents 
 of the magnetic and electronic chains.\cite{white-1996} We obtain
\begin{eqnarray}
\mathcal{H}_s&=&\mathcal{H}_0+\mathcal{H}_{int}\nonumber\\
\mathcal{H}_0&=& \frac{2\pi v_1}{3}(\colon \mathbf{J}_{R,1} \cdot \mathbf{J}_{R,1} +\colon \mathbf{J}_{L,1} \cdot \mathbf{J}_{L,1}\colon) \nonumber\\ 
&+&\frac{2\pi v_2}{3}(\colon \mathbf{J}_{R,2} \cdot \mathbf{J}_{R,2} \colon +\colon \mathbf{J}_{L,2} \cdot \mathbf{J}_{L,2}\colon ) \label{Hsfree}\nonumber\\
\mathcal{H}_{int}&=&  g_1 \mathbf{J}_{R,1}\cdot \mathbf{J}_{L,1}+g_2 \mathbf{J}_{R,2}\cdot \mathbf{J}_{L,2}
\nonumber\\&+&g_3( \mathbf{J}_{R,1}\cdot \mathbf{J}_{L,2}+\mathbf{J}_{L,1}\cdot \mathbf{J}_{R,2})\nonumber\\
&+&g_4 (\mathbf{J}_{R,1}\cdot \mathbf{J}_{R,2}+\mathbf{J}_{L,1}\cdot \mathbf{J}_{L,2})
\label{HHHKcurr}
\end{eqnarray}
where $1$ and $2$ label the electronic system and the spin chain, respectively, and 
$v_1\sim 2 a t$ and $v_2 \sim a J_1$. All the operators that we have included are marginal, and they are marginally relevant for $g>0$ and marginally irrelevant otherwise. 
Notice that in the effective interaction of Eq. \eqref{HHHKcurr} we have not included a possible coupling between the spin-density wave order parameters 
of the 1D electronic system and of the Heisenberg spin chain. 
For general filling of the 1D electronic system this interaction is not allowed (or, rather it is strongly irrelevant) since the two systems have different ordering wave vectors. 
However it is a relevant perturbation in the case of the Kondo insulator where it plays a key role.

The coupling constant $g_2$ parametrizes the strength of the backscattering term of the magnetic chain. 
According to the discussion of Section \ref{sec:model} this coupling is irrelevant in the absence of the next-nearest-neighbor (frustrating) exchange interaction $J_2$ 
and should became relevant at some critical value $J_{2c}$, past which the spin chain becomes dimerized and has a spin gap. 
Thus, the bare value of the coupling constant  $g_2$ should change sign at $J_{2c}$, and close to this critical point it should have the simple form 
$g_{2}\sim a (J_2-J_{2c})$. This will be the initial value for our renormalization group (RG) analysis. 
Instead, the backscattering interaction of the electronic chain is marginally irrelevant (for repulsive microscopic interactions) and hence the bare value of $g_1$ is negative. 
Thus, although the Hamiltonian for the spin sectors of the two subsystems have the same form, they are not equivalent.

 In Eq. \eqref{HHHKcurr} the  Kondo term has been split into the $g_3$ term  for the coupling of currents with the different chirality, and  the $g_4$ term for the 
 coupling of currents of the same chirality. Their  initial (bare) values  are $g_{30}=g_{40}\sim a J_K$. Finally $g_1$ 
 corresponds to a possible backscattering term in the electronic chain induced by electronic correlations. For repulsive interactions it has a negative bare value, $g_{10}<0$. 

The (chiral) spin currents of the electronic system and of the spin chain generate, separately, an $SU(2)_1$ Kac-Moody algebra.  
Consequently, the chiral spin currents $\mathbf{J}_{aR,L}$ (with $a=1,2$) have the operator product expansion (OPE)\cite{difrancesco-1997}
\begin{align}
J^{\alpha}_{L}(z_a) J^{\beta}_{L}(w_b)
\sim&\frac{\delta_{ab} \delta^{\alpha \beta}}{8 \pi(z_a-w_b)^2 }+\sum_{\gamma}\frac{\delta_{ab}\epsilon^{\alpha \beta \gamma}
J^{\gamma}_L(w_{b})}{2\pi(z_a-w_b)}
\nonumber \\ 
J^{\alpha}_{R}(\bar{z}_a) J^{\beta}_{R}(\bar{w}_b)
\sim&\frac{\delta_{ab} \delta^{\alpha \beta}}{8 \pi(\bar{z}_a-\bar{w}_b)^2 }+\sum_{\gamma}\frac{\delta_{ab}\epsilon^{\alpha \beta \gamma} 
J^{\gamma}_L(\bar{w}_{b})}{2\pi(\bar{z}_a-\bar{w}_b)}
\nonumber\\
\label{OPESJJ}
\end{align}
where $\alpha,\beta=x,y,z$; $a,b=1,2$ and $z_a=i x +v_a \tau$, where $\tau=it$ is the imaginary time. 

From Eq. \eqref{OPESJJ} we can obtain the one-loop RG equations using the procedure outlined in the Appendix A of Ref. \onlinecite{balents-1996} 
(or the general approach described in Refs. \onlinecite{cardy-book,fradkin-2013}.)
It easy to see that $g_4$-term does not contribute to the lowest order RG equations. The remaining equations are thus decoupled and read as follows
\begin{equation}
\frac{dg_1}{dl}=\frac{g_1^2}{2 \pi v_1}, \qquad \frac{dg_2}{dl}=\frac{g_2^2}{2 \pi v_2}, \qquad
 \frac{dg_{3}}{dl}=\frac{g_3^2}{\pi (v_1+v_2)}
\label{RGKH}
\end{equation}  
Thus, starting from a negative value $g_1$ (slowly) approaches zero, and will be neglected in the sequel. 
For next-nearest-neighbor (frustrating) exchange coupling $J_2$ smaller than the critical value $g_{20}<0$, and $g_2$ is also (marginally) irrelevant. 
In this regime the  flow is controled by the marginally relevant Kondo coupling, and the system flows to the same fixed point of the unfrustrated Kondo-Heisenberg chain. 
In this regime the system is in a PDW phase of the unfrustrated chain, as discussed in Ref. [\onlinecite{berg-2010}] 
(using results from Refs. [\onlinecite{zachar-1996,zachar-2001,zachar-2001b}].)

We now follow the results of White and Affleck\cite{white-1996} to analyze the situation for $J_2>J_{2c}$. In this regime  the bare value of $g_2$ is positive, $g_{20}>0$, 
and the backscattering interaction of the spin chain is marginally relevant.   We thus find two stable phases depending on which coupling $g_2$ or $g_3$ 
reaches first its strong coupling limit under the RG flow of Eq.\eqref{RGKH}.
If $g_2$ wins,  the fixed point describes a dimerized spin chain (with a spin gap) and a decoupled electronic system. In this limit the Kondo coupling is irrelevant. 
Conversely, if $g_3$ reaches the strong coupling limit first, the system is again in the PDW phase in which dimerization is suppressed. 

We can see how these phases arise  using abelian bosonization of the $SU(2)_1$ Kac-Moody algebra.\cite{difrancesco-1997} 
In this case the chiral spin currents can be represented in terms of the chiral bosonic fields $\varphi_{sa}$ (with $s=R,L$ and $a=1,2$ representing each Heisenberg chain)
\begin{equation}
J_{s a}^{\pm}=\frac{1}{2\pi a} e^{\mp i \sqrt{8\pi}\varphi_{s a}}, \qquad
J^z_{s a}=\frac{1}{\sqrt{2\pi}}\partial_x \varphi_{s a}
\label{SU(2)1}
\end{equation}    
These chiral currents have (as they should) scaling dimension $1$.
In this representation the spin chain backscattering interaction, the $g_2$ term in Eq. \eqref{HHHKcurr}, becomes
\begin{eqnarray}
g_2\mathbf{J}_{R 2}\cdot\mathbf{J}_{L 2}&=&\frac{g_2}{(2\pi a)^2}\cos(\sqrt{8\pi}\varphi_2)+\frac{g_2}{2\pi}\partial_x \varphi_{R 2} \partial_x \varphi_{L 2}\nonumber\\
&&
\label{g2boso}
\end{eqnarray}
where $\varphi_a=\varphi_{R a}+\varphi_{L a}$, and $\theta_a=\varphi_{L a}-\varphi_{R a}$ is its dual field.

In the regime in  which  the RG flow drives $g_2$ to strong coupling, the field $\varphi_2$ is pinned at the minimum of the cosine, 
i.e. at the values $\varphi_{2}=(n+\frac{1}{2}) \sqrt{\frac{\pi}{2}}$. In this phase the order parameter $\epsilon$ defined in Eq.\eqref{eq:dimerization}, 
has a non-vanishing expectation value, and the system dimerizes (the $D$ phase in what follows) breaking the translational symmetry 
and has a finite spin gap in its spectrum. 
Since the magnetic chain has a spin gap, its spin-spin correlation functions are short ranged. 
However, White and Affleck\cite{white-1996} found that in the dimerized phase the spins of the magnetic chain have short-ranged incommensurate antiferromagnetic order. 
On the other hand, the electronic chain in this phase remains decoupled with gapless charge and magnetic excitations.

In the opposite case,   in which $g_3$ flows to strong coupling, the system flows to the pair-density-wave fixed point of the unfrustrated Kondo-Heisenberg chain.
Let us summarize, for completeness, how this happens. The inter-chain backscattering operators have the bosonized expression
\begin{eqnarray}
\mathbf{J}_{R 1}\cdot\mathbf{J}_{L 2}+\mathbf{J}_{L 1}\cdot\mathbf{J}_{R 2}&=&\nonumber\\
\frac{1}{(2\pi a)^2}\Bigg\{\cos[\sqrt{8\pi}(\varphi_{R 1}+\varphi_{L 2})]&+&
\cos[\sqrt{8\pi}(\varphi_{L 1}+\varphi_{R 2})]\Bigg\}+\nonumber\\
\frac{1}{2\pi}\Bigg[\partial_x \varphi_{R 1} \partial_x \varphi_{L 2}&+&\partial_x \varphi_{L 1} \partial_x \varphi_{R 2}\Bigg]\nonumber\\
\label{g3boso}
\end{eqnarray}
The first term in Eq.\eqref{g3boso} can be written in the form
\begin{eqnarray}
\frac{1}{(2\pi a)^2}\bigg\{\cos[2\sqrt{\pi}(\varphi_{+}+\theta_{-})]+
\cos[2\sqrt{\pi}(\varphi_{+}-\theta_{-})]\bigg\}\nonumber\\
\label{eq:relevant-pdw}
\end{eqnarray}
where $\varphi_{\pm}=\frac{1}{\sqrt{2}} \left(\varphi_{2}\pm\varphi_{1}\right)$ and $\theta_{\pm}=\frac{1}{\sqrt{2}} \left(\theta_{2}\pm\theta_{1}\right)$.

The operator shown in Eq.\eqref{eq:relevant-pdw} is the marginally relevant interaction found in Ref. [\onlinecite{zachar-2001}]. 
Its presence in the effective action drives the system to a fixed point in which the fields 
$2\sqrt{\pi}\varphi_+$ and $2\sqrt{\pi}\theta_-$ are pinned at the values 
$2n\pi$ and $(2n+1)\pi$, or $(2n+1)\pi$ and $2n\pi$. The corresponding phase was analyzed in  Refs. [\onlinecite{zachar-2001}] and [\onlinecite{berg-2010}] 
where it was shown that it has gap for all spin excitations.
In this phase there still is a decoupled gapless charge sector. 

A remarkable feature of this phase is that the only order parameters that exhibit quasi-long-range order 
(i.e. have power-law correlations) are composite operators made from observables of the electronic chain and the spin chain that, separately, have short-range correlations since their correlation functions fall-off exponentially fast with distance. 
In particular this phase is characterized by the pair-density wave 
order  parameter, 
\begin{equation}   
\Delta_{PDW}=\bm{\Delta}_{TS}\cdot\bm{N}_2
\label{OPDWdef}
\end{equation}
where $\bm{\Delta}_{TS}=i\sum_{\alpha,\beta}\psi_{\alpha R,1}(x)(\bm{\sigma}\sigma_y)_{\alpha,\beta}\psi_{\sigma L,1}$ is the spin triplet pairing operator of the electronic 
system (``chain $1$''), and $\mathbf{N}_2$ is the N\'eel order parameter of the spin chain (``chain $2$''). 
Thus, the PDW order parameter is a four fermion operator. The PDW operator of Eq. \eqref{OPDWdef} has the bosonized form
 \begin{eqnarray}   
\Delta_{PDW}&=&\frac{ e^{-i\sqrt{2\pi}\theta_c}}{2(\pi a)^2}[2 \cos(\sqrt{4 \pi} \theta_-)+\cos(\sqrt{4 \pi} \varphi_+)\nonumber\\
&-&\cos(\sqrt{4 \pi} \varphi_-)]
\label{OPDWboso}
\end{eqnarray}
where we have omitted an oscillatory factor with wave vector $Q_{PDW}=\pi$. 
From this expression it is apparent that in the PDW phase the operators in brackets in Eq.\eqref{OPDWboso} have finite (and non-vanishing) expectation values and, 
as a result, the PDW order parameter has power-law correlations of the form
\begin{eqnarray}
\langle \Delta_{PDW}(x)\Delta^\dagger_{PDW}(0) \rangle \sim\frac{1}{x^{\eta_{PDW}}}
\end{eqnarray}
The exponent takes the value $\eta_{PDW}=1$ if the electronic chain is non-interacting, while for repulsive interactions (with $K_c>1$) it increases to the value  
$\eta_{PDW}=K_c>1$.

Berg {\it et al.}\cite{berg-2009b} showed that an ordered PDW state with ordering wave vector $Q$ always has a subleading uniform 
superconducting order but with charge $4e$ instead of $2e$ as in a conventional superconductor. 
In Ref.[\onlinecite{berg-2010}] showed that the spin-gap phase of the Kondo-Heisenberg chain has PDW quasi-long-range order as well as charge
 $4e$ quasi-long-range order (albeit with a larger critical exponent). For a PDW with ordering wave vector $Q=\pi$ (as in the present case), 
 the charge-$4e$ superconducting order parameter $\Delta_{4e}$ is simply the square of the PDW order parameter. Hence we can make the identification 
 \begin{equation}
 \Delta_{4e} \sim \textrm{const.} \times e^{i2\sqrt{2\pi}\theta_c}
 \label{Delta4e}
 \end{equation}
 This order parameters has scaling dimension $2K_c$, and its correlation function falls-off with an 
 exponent $\eta_{4e}=4K_c$.

The RG flows of Eq. \eqref{RGKH} determine the structure of the phase diagram in  this weak coupling regime. 
The RG flows are marginally unstable on both coupling constants and hence the effective couplings run to their strong coupling regime. 
As usual, the velocities are only affected by irrelevant operators and acquire at most a finite renormalization. 
Hence the velocities do not affect the RG flow of the dimensionless coupling constants $g_2$ and $g_3$ since $g_1$ is a marginally irrelevant coupling.
The dimensionless coupling constants have initial values $g_{20}\sim (J_2-J_{2c})/t$ and $g_3\sim J_K/t$.

At the level of the one-loop RG of Eq.\eqref{RGKH}, which is accurate for $g_2$ and $g_3$ small, the location of the phase boundary (the separatrix of the RG flow) 
between PDW and the dimerized phases is the straight line $g_3=\frac{g_2}{2} \left(1+\frac{v_1}{v_2}\right)$ or, equivalently in terms of the macroscopic parameters, 
$J_K=\frac{1}{2}(J_2-J_{2c})\left(1+\frac{2t}{J_1}\right)$. Above this phase boundary the system is in the PDW phase, 
and below this phase boundary it is in the dimerized phase. Along the phase boundary both coupling constants $g_2$ and $g_3$ flow to strong coupling at the same rate. 
Thus, the quantum phase transition between the  PDW phase and the dimerized phase is likely to be first order. 

We conclude that quantum frustration of the spin chain leads to irrelevant and essentially unobservable effects in the PDW phase. 
On the other hand quantum frustration leads to a dimerized phase and the Kondo coupling has essentially irrelevant effects in this phase. 
This is so because in this phase there is a spin gap in the (frustrated) spin chain producting an effective decoupling from the electronic chain. 
It is well known (and easy to see) that if the spin chain is treated classically, its ground state is an incommensurate spiral. 
The numerical (DMRG) calculations of White and Affleck show that there is short-range incommensurate order inside the dimerized phase.

%boundary can be found by computing the scales $l^*$ and $l^{\prime *}$ such $\bar{g}_2(\l^*)=1$ and $\bar{g}_3(\l^{\prime *})=1$ are equal. 
%Solving Eq. (\ref{RGKH}) and imposing this condition we obtain:
%\begin{eqnarray}
%\bar{g}_{30}=\frac{1}{\frac{2\bar{v}_1}{(\bar{v}_1+1)}\left(\frac{1}{\bar{g}_{10}}-1\right)+1}
%\end{eqnarray}
%\begin{figure}[hbt]
%\centering
%\includegraphics[width=0.4\textwidth]{phdiagj1j2jkwf.eps}
%\caption{
%Schematic phase diagram  $\bar{g}_{30}\sim\frac{J_K}{t}$ vs $\bar{g}_{20}\sim\frac{(J_2-J_{2c})}{t}$ for different values of $\bar{v}_1\sim\frac{J_1}{t}$.}
%\label{phdiagwf}
%\vspace{0.1cm}
%\end{figure}
%In Fig. \ref{phdiagwf} we show our schematic phase diagram in the weak frustrations limit. We see that for increasing values of $\frac{J_1}{t}$ the 
%relative zone of the D phase decrease. This is natural because the relative frustration $\frac{J_2-J_{2c}}{J_1}$ weakens with increasing this value.     

\section{The $J_2 \gg J_1$ regime}
\label{sec:J2ggJ1}

We now turn to the limit in which the next-nearest-neighbor exchange interaction $J_2$ is strongest than $J_1$. In this regime the treatment used in the preceding 
section is not adequate. Instead, it is convenient to think the magnetic chain as  a zig-zag two-leg ladder, i.e. two spin chains with nearest neighbor exchange 
$J_2$, and weakly coupled by a inter-chain (zig-zag) exchange interaction $J_1$. In this picture the frustrated spin chain is a triangular (``trestle") ladder. 
In the extreme case  $J_1=0$ the system we are considering becomes two Heisenberg antiferromagnetic chains coupled with an electronic chain by the Kondo 
exchange interaction. 
In the limit in which  the Kondo coupling is also weak, the starting point of our treatment consists of three chains (two magnetic chains and an interacting electronic chain) weakly 
coupled by $J_K$ and $J_1$. This regime turns out to be quite rich.

The zig-zag antiferromagnetic ladder in the limit $J_2 \gg J_1$ was considered by White and Affleck\cite{white-1996} who treated the system as two weakly 
coupled Heisenberg chains. In this limit, the effective low energy theory of this zig-zag ladder includes a marginally relevant backscattering coupling between the 
SU(2) spin currents of the two chains. However, due to the special symmetry of this ladder the (otherwise relevant) inter-chain coupling of their 
N\'eel order parameters is absent. 
White and Affleck also showed that the marginally relevant inter-chain backscattering interaction causes the RG to flow to a strong coupling fixed point with finite 
(albeit exponentially small) dimerization and an also exponentially small spin gap. 
These results were confirmed using a DMRG. Furthermore in this regime the zig-zag ladder has short-range incommensurate spiral order. 
Thus, for the problem of interest here, we conclude that if the Kondo exchange interaction is also very small, the low energy physics is that of a dimerized 
zig-zag chain essentially decoupled from a one-dimensional Luttinger liquid with gapless and decoupled charge and spin excitations. 
This is the same phase that we encountered in the preceding section.

Another limit of interest is the case in which  $J_1\to 0$ while holding $J_K$ finite (and small), in which the system reduces to two antiferromagnetic chains 
and  a conducting fermionic chain coupled by a Kondo exchange interaction $J_K$.  Up to some simple redefinitions, in this limit the problem we are interested in 
is related to the problem of over-screened (or multi-channel) Kondo-Heisenberg chains considered by Azaria and Lecheminant,\cite{azaria-2000} 
(who also considered the frustrated three-leg Heisenberg ladder.) We will see below that the spin sector of the problem we are interested in is equivalent, in the limit 
$J_1 \to 0$, to the spin sector of an over-screened Kondo-Heisenberg model. 

The main result of the work by Azaria and Lecheminant is that the Kondo exchange interaction, which is marginally relevant at $J_2=0$, 
drives the spin sector to a finite infrared-stable fixed point with non-trivial properties.  This non-trivial fixed point is a one-dimensional analog of the 
multi-channel Kondo problem. 
At this finite fixed point the spin sector decouples into two ``chirally stabilized'' spin liquids, first discussed by Andrei, Douglas and Jerez.\cite{andrei-1998} 
We will use the exact solution of Azaria and Lecheminant for the two-channel case, which is relevant to our problem, and discuss the effects of the coupling $J_1$ 
at this non-trivial fixed point. 

To see how this works let us consider the effective (bosonized) low energy field theory of the frustrated Kondo-Heisenberg chain in the limit $J_2\gg J_1$ and $J_2\gg J_K$.
In the low energy limit the frustrated Kondo-heisenberg chain system exhibits spin-charge separation. 
The total effective low energy Hamiltonian density $\mathcal{H}=\mathcal{H}_c+\mathcal{H}_s$  is a sum of the Hamiltonian  $\mathcal{H}_c$ for the charge sector 
and  $\mathcal{H}_s$ for the spin sector.

The charge sector is described, as in the preceding section, by the charge Bose field $\phi_c$ and its dual field $\theta_c$,  with a fixed charge Luttinger parameter 
$K_c$ and a velocity $v_c$. 
The effective Hamiltonian for the charge sector, $\mathcal{H}_c$, describes the charge degrees of freedom of the fermionic system and, hence, 
is the same as the one given in Eq.\eqref{Hcharge}. 

In this regime, the spin sector is described by the chiral spin currents of the two weakly coupled legs of the zig-zag ladder and the chiral spin currents 
of the 1D electronic system. The Hamiltonian density $\mathcal{H}_s$ of the spin sector is given by
\begin{align}
\mathcal{H}_s=&\mathcal{H}_0+\mathcal{H}_{\rm int}\nonumber\\
\mathcal{H}_0=& \frac{2\pi v_1}{3}\bigg(\colon \mathbf{J}_{R,1} \cdot \mathbf{J}_{R,1} +\colon \mathbf{J}_{L,1} \cdot \mathbf{J}_{L,1}\colon 
\nonumber\\
&+\colon \mathbf{J}_{R,2} \cdot \mathbf{J}_{R,2}\colon +\colon \mathbf{J}_{L,2} \cdot \mathbf{J}_{L,2}\colon \bigg)
\nonumber\\ 
&+
\frac{2\pi v_2}{3}\bigg(\colon \mathbf{J}_{R,3} \cdot \mathbf{J}_{R,3} \colon +\colon \mathbf{J}_{L,3} \cdot \mathbf{J}_{L,3}\colon \bigg) 
\label{H03chain}\\
\mathcal{H}_{\rm int}=& g_1 \bigg[\left(\mathbf{J}_{R,1}+\mathbf{J}_{R,2}\right)\cdot \mathbf{J}_{L,3}+\left(\mathbf{J}_{L,1}+\mathbf{J}_{L,2}\right)\cdot 
\mathbf{J}_{R,3}\bigg]
\nonumber\\
+&g_2 \left(\mathbf{J}_{R,1}\cdot \mathbf{J}_{L,2}+\mathbf{J}_{R,2}\cdot \mathbf{J}_{L,1}\right)
\nonumber\\
+& {\bar g} \left( \mathbf{J}_{R,1} \cdot \mathbf{J}_{L,1}+\mathbf{J}_{R,2} \cdot \mathbf{J}_{L,2}\right)+{\bar g}' \; \mathbf{J}_{R,3} \cdot \mathbf{J}_{L,3}
\nonumber\\
&
\label{HHKSF}
\end{align}
Here $1$ and $2$  label the SU(2) spin currents the two magnetic chains and $3$ labels the SU(2) spin current of the electronic chain. 
The bare values of the spin velocities, $v_1$ and $v_2$, and of the effective coupling constants $g_1$ and $g_2$, are respectively given by  
$v_1\simeq a J_2$, $v_2\simeq a t$, $g_{1}\simeq a J_K$, $g_{2}\simeq a J_1$. 
The last line in Eq.\eqref{HHKSF} represents the marginally irrelevant backscattering intra-interaction of the spin currents of the two spin chains. 
The effective coupling constants $\bar g$ and $\bar g'$ are given by:   $\bar g\sim -J_1<0$ for the spin chain,
and $\bar g'\approx -2U/t <0$ for the 1D electronic system  (where $U$ and $t$ are the on-site Hubbard interaction and $t$ is the nearest-neighbor hopping amplitude), 
respectively.
As in Section \ref{sec:J2llJ1}, in Eq. \eqref{HHKSF} we have neglected the redundant term that couples the spin currents with the same chirality, 
which leads to a finite renormalization of the velocities. 

Also as in Section \ref{sec:J2llJ1}, 
here too we  have  not included in Eq. \eqref{HHKSF} a coupling between the spin-density wave order parameter of the 1D electronic system 
and the N\'eel order parameters of the weakly coupled spin chains since their associated ordering wave vectors are different and hence the coupling 
is not allowed in a system with translation invariance. Similarly, we have not included the (relevant) coupling between the N\'eel order parameters 
of the two weakly coupled spin chains since it is forbidden by the symmetries of the zig-zag chain. 
The absence of these (potentially most strongly relevant) interactions has important consequences for the stability analysis of the fixed points
 and for the structure of the phase diagram. 
 
 The representation of the degrees of freedom of the frustrated quantum Heisenberg chain in terms of a zig-zag chain in the  effective Hamiltonian of Eq.\eqref{HHKSF} is invariant under the exchange of the two spin chains. This parity symmetry is actually broken explicitly in the lattice model. 
Nersesyan Gogolin and Essler\cite{nersesyan-1998} have shown that in order to account for these parity-breaking effects it is necessary to include in the effective Hamiltonian of Eq.\eqref{HHKSF} operators of the form $\textbf{N}_1 \cdot \partial_x \textbf{N}_2-\textbf{N}_2\cdot \partial_x \textbf{N}_1$. However they also showed (and was more recently confirmed in numerical calculations\cite{hikihara-2010}) that the effects of these parity-breaking terms are suppressed unless there is a large (easy plane) magnetic anisotropy and hence do not contribute in the SU(2)-invariant system. Operators with a similar structure mixing the SDW order parameter of the electronic chain and the N\'eel order partners of the spin chains are not allowed by the mismatch of the ordering wave-vectors (which renders them strongly irrelevant). Nevertheless we find that there is an instability out of the phase governed by the Toulouse point (discussed below) triggered by operators that break parity. These operators can be regarded as a remnant of the parity-breaking effects discussed by Nersesyan, Gogolin and Essler.

The effective field theory for the spin sector described by Eqs. \eqref{H03chain} and \eqref{HHKSF} has three SU(2)$_1$ spin currents 
(two SU(2)s for each spin chain and one SU(2)  for the electronic system). Thus it is a perturbed SU(2)$_1 \times $SU(2)$_1 \times $SU(2)$_1$ conformal field theory.  
In addition the system has a $U(1)$ charge sector which (in the low energy limit) is decoupled from the spin sector. 
 In the absence of the perturbations of Eq.\eqref{HHKSF} the spin sector has central charge $c=3$. 
 
 \subsection{$J_K\ll J_1\ll J_2$: Dimerized Phase}
\label{sec:JKllJ1llJ2}

 The effect of the perturbations is to open up gaps thus driving the system to a fixed point with a smaller central charge.
The RG equations for the effective field theory of Eq.\eqref{HHKSF} have the same decoupled form as before, i.e.
\begin{equation}
\frac{dg_1}{dl}=\frac{g_1^2}{\pi (v_1+v_2) }, \qquad
 \frac{dg_{2}}{dl}=\frac{g_2^2}{2\pi v_1}
\label{RGKHSF}
\end{equation}  
The RG flows are decoupled, as in the case of section \ref{sec:J2llJ1}. Which gap opens up depends on the relative strengths of the perturbations.

For $J_K \ll J_1$, $g_2 \gg g_1$,  the RG flows to the strong coupling fixed-point (of $g_2$) of the dimerized zig-zag chain which is now gapped. 
Since the perturbation with coupling constant $g_2$ is marginally relevant the gap is exponentially small (up to logarithmic corrections due to the 
marginally irrelevant coupling $\bar g$).\cite{white-1996}
In this phase the 1D electronic system and the zig-zag chain are decoupled at low energies. 
This is the same dimerized phase we found in Section \ref{sec:J2llJ1} 
with the only difference that when $J_2\gg J_1$ the zig-zag chain has short-range incommensurate spiral order.\cite{white-1996} 
Thus, in this phase we have a gapless charge mode and a gapless spin mode, both belonging to the decoupled electronic system.

On the other hand, for $J_K \gg J_1$ the RG flows to different strong coupling regime that will be described below. 
From the results of Section \ref{sec:J2llJ1} we know that one possibility is that this phase may also be the PDW phase that arises for $J_K \gg J_1$ (and $J_2\ll J_1$). 
The other possibility is that the strong frustration regime is in a different phase altogether. 

The phase boundary that separates the dimerized phase from the non-trivial phase is the separatrix of the RG flow
 of Eq.\eqref{RGKHSF}. This flow has a separatrix at $g_1=\frac{g_2}{2}\left(1+\frac{v_2}{v_1}\right)$ or, equivalently, 
 $\frac{J_K}{J_2}=\frac{1}{2} \frac{J_1}{J_2} \left(1+\frac{t}{J_2}\right)$.

%transition line between the phase dominated by ${g_1}$ and the one by ${g_3}$ could be determined by the same procedure used in the previous Section. 
%We obtain:
%\begin{eqnarray}
%\bar{g}_{10}=\frac{1}{\frac{2\bar{v}_1}{(\bar{v}_1+1)}\left(\frac{1}{\bar{g}_{20}}-1\right)+1}
%\end{eqnarray}
%In Fig. \ref{phdiagsf} we show our schematic phase diagram in the strong frustrations limit. 
%As discussed in Ref. \onlinecite{white-1996} when $g_1$ increases a spin gap open in the two chain
%and a dimerization take places in the interchain coupling. This is the same phase D we have found in weak coupling limit.

%\begin{figure}[hbt]
%%\centering
%\includegraphics[width=0.4\textwidth]{phdiagj1j2jksf.eps}
%\caption{
%Schematic phase diagram in the strong frustration limit. We show the transition line in the plane $\bar{g}_{10}\sim\frac{J_K}{t}$ vs $\bar{v}_{1}\sim\frac{J_2}{t}$ 
%for different values of 
%$\bar{g}_{20}\sim\frac{J_1}{t}$.}
%\label{phdiagsf}
%%\vspace{0.1cm}
%\end{figure}
%In the opposite situation, when $g_2(l)\rightarrow\infty$  we expect as in the case of the one spin chain coupled with a 1DEG, the development of a 
%spin gap with PDW dominant correlations. 

\subsection{The $J_1\ll J_K \ll J_2$ regime}
\label{sec:J1JKJ2}

The nature of the phase in the $J_1 \ll J_K\ll J_2$ regime cannot be accessed by the perturbative RG of Eq.\eqref{RGKHSF} and must be determined non-perturbatively. This regime is governed by a finite, infrared stable,  fixed point that was first discussed by Azaria and Lecheminant\cite{azaria-2000,azaria-2000b} 
(who were interested in a somewhat different problem) who constructed a non-trivial fixed point of a type first discussed by Andrei, Douglas, and Jerez\cite{andrei-1998} using a non-perturbative approach.
We will consider first the case $J_1=0$ and use the results of Refs. [\onlinecite{azaria-2000}] and [\onlinecite{azaria-2000b}] 
to construct the non-trivial fixed point at a finite value of the Kondo coupling constant. 
Once this fixed point is identified we will assess the role of the perturbation with coupling constant $J_1$ at the non-trivial fixed point in order 
to determine the actual nature of the phase. In particular if the fixed point is unstable even for small $J_1$ we will see that the resulting phase has the 
same properties as the pair-density-wave phase of the unfrustrated Kondo-Heisenberg chain. 
However if the non-trivial fixed point is stable then there is a new phase (for weak enough $J_1$) which is distinct from the PDW phase and from the dimerized phase. 

\subsubsection{$J_1=0$: Non-trivial Finite Fixed Point}

We begin by considering first the case $J_1=0$ and summarize the construction of the nontrivial finite fixed point by Azaria and Lecheminant.\cite{azaria-2000} 
So, we will consider the spin sector of our system with Hamiltonian $\mathcal{H}_s$, see Eqs.\eqref{H03chain} and \eqref{HHKSF}, 
and set for now $g_2=\bar g=\bar g'=0$. The Hamiltonian density now reads
\begin{align}
\mathcal{H}_s=&\frac{2\pi v_1}{3} \sum_{i=1,2} \Big(: \mathbf{J}_{i,R} \cdot \mathbf{J}_{i,R}:+: \mathbf{J}_{i,L} \cdot \mathbf{J}_{i,L}:\Big)\nonumber\\
+&\frac{2\pi v_2}{3} \Big(: \mathbf{J}_{3,R} \cdot \mathbf{J}_{3,R}:+: \mathbf{J}_{3,L} \cdot \mathbf{J}_{3,L}:\Big) \nonumber\\
+&g_1 \Big(\sum_{i=1,2} \mathbf{J}_{i,R} \cdot \mathbf{J}_{3,L}+\sum_{i=1,2} \mathbf{J}_{i,L} \cdot \mathbf{J}_{3,R}\Big)
\label{Heff-g1}
\end{align}
The free part of this Hamiltonian has an SU(2)$_1 \times $ SU(2)$_1 \times $ SU(2)$_1$ symmetry which is partially broken by the interaction term. 
Since the interaction  retains the symmetry of the exchange of the spin sectors of the two spin chains (here denoted by $1$ and $2$), 
it is natural to rewrite this system in terms of the total chiral  right- and left-moving chiral currents current $\mathbf{I}_{R,L}=\mathbf{J}_{1,R}+\mathbf{J}_{2,R}$.
The chiral currents $\mathbf{I}_{R,L}$ are the generators of  two chiral SU(2)$)_2$ Kac-Moody algebras (one for each chirality). 

 We will first rewrite the two decoupled chains in terms of a theory of four Majorana fermions.\cite{allen-1997,shelton-1996} 
 This is possible since a conformal field theory with current algebra SU(2)$_1 \times $ SU(2)$_1$ is equivalent to a conformal field theory of the 
 Wess-Zumino-Witten model with current algebra SO(4)$_1$. This model has a well known representation in terms of four free Majorana fermions\cite{witten-1984}
\begin{align}
\frac{2\pi v_1}{3}& \sum_{i=1,2} \Big(: \mathbf{J}_{i,R} \cdot \mathbf{J}_{i,R}:+: \mathbf{J}_{i,L} \cdot \mathbf{J}_{i,L}:\Big)
\nonumber\\
&=-\frac{v_1}{2}\sum_{i=0,1,2,3} \Big(\xi_R^i i \partial_x \xi_R^i-\xi_L^i i\partial_x \xi_L^i\Big)
\label{SO(4)}
\end{align}
where $\xi_{s}^i(x)$ (with $i=0,1,2,3$) denotes the four species of Majorana fermions with both chiralities ($s=R, L$), and satisfy canonical anti-commutations relations, 
$\{ \xi_{s}^i(x),\xi_{s'}^j\}=\delta_{ij} \delta(x-y)$  and $\{ \xi_R^i(x), \xi_L^j(y)\}=0$.

 In terms of the Majorana fields, the chiral  SU(2)$_2$ currents, $I_s^a$  (with $a=1,2,3$ and chirality $s=R,L$), which are  three of the six chiral  currents of SO(4)$_1$ 
 (with both chiralities), are given by
\begin{equation}
I_{s}^a=J_{1,s}^a+J_{2,s}^a=-\frac{i}{2} \epsilon^{abc}\xi_s^b\xi_s^c
\label{SU2_2}
\end{equation}
The remaining three chiral currents  of SO(4)$_1$ are (with $s=R,L$)
\begin{equation}
K_s^a=i\xi^a_s\xi^0_s
\label{SO4}
\end{equation}
where $\mathbf{K}_{R,L}=\mathbf{J}_{1,R,L}-\mathbf{J}_{2,R,L}$.
This decomposition allows us to rewrite the two decoupled spin chains in terms of the SU(2)$_2$ chiral currents. 
It is equivalent to the identification \cite{goddard-1986,ginsparg-1988,difrancesco-1997} SU(2)$_1 \times$ SU(2)$_1 \simeq $ SU(2)$_2 \times \mathbb{Z}_2$, 
where the $\mathbb{Z}_2$ factor will be represented by a critical Ising model, i.e. a massless Majorana fermion which cannot be written in the Sugawara form 
(as a quadratic form in chiral currents). 

Following Ref. [\onlinecite{azaria-2000}] we now rewrite the Hamiltonian density of  Eq.\eqref{Heff-g1} in the form
\begin{equation}
\mathcal{H}_s=- \frac{v_1}{2} \Big(\xi_R^0 i \partial_x \xi_R^0-\xi_L^0 i \partial_x \xi_L^0\Big)+\mathcal{H}_1+\mathcal{H}_2
\label{SU(2)2+Z2}
\end{equation}
where
\begin{align}
\mathcal{H}_1=& \frac{\pi v_1}{2} :\mathbf{I}_R \cdot \mathbf{I}_R:+ \frac{2\pi v_2}{3} :\mathbf{J}_{3,L} \cdot \mathbf{J}_{3,L}:+ g_1 \mathbf{I}_R \cdot \mathbf{J}_{3,L}
\nonumber\\
\mathcal{H}_2=& \frac{\pi v_1}{2} :\mathbf{I}_L \cdot \mathbf{I}_L:+ \frac{2\pi v_2}{3} :\mathbf{J}_{3,R} \cdot \mathbf{J}_{3,R}:+ g_1 \mathbf{I}_L \cdot \mathbf{J}_{3,R}
\nonumber\\
&
\label{H1H2}
\end{align}
Obviously, $\mathcal{H}_1$ and $\mathcal{H}_2$ commute with each other, $[ \mathcal{H}_1,\mathcal{H}_2]=0$.
The Hamiltonians $\mathcal{H}_1$ and $\mathcal{H}_2$ involve the chiral SU(2)$_2$ currents $\mathbf{I}_{R,L}$ 
and the chiral SU(2)$_1$ spin currents $\mathbf{J}_{3,R,L}$ of the electronic system. 
In addition, these two Hamiltonians also commute with the free Majorana Hamiltonian shown in Eq.\eqref{SU(2)2+Z2}, 
which represents a $\mathbb{Z}_2$ (Ising) sector. 
The Hamiltonians $\mathcal{H}_1$ and $\mathcal{H}_2$ have the form of the chirally-stabilized spin liquid theory of 
Andrei, Douglas and Jerez\cite{andrei-1998} who showed that each describes a theory controlled by a finite infrared-stable fixed point. 
Although the fixed point of the theory of Hamiltonian $\mathcal{H}_1$ is not chirally-symmetric, the sum $\mathcal{H}_1+\mathcal{H}_2$ is chirally-symmetric.

The most direct way to identify the finite non-trivial fixed point is to use abelian bosonization approach of Azaria and Lecheminant\cite{azaria-2000} 
which solves the problem by  formally breaking  the SU(2) symmetry  down to a U(1) subgroup. In order to do this we will formally break the SU(2) 
symmetry of the interaction terms of  the Hamiltonians $\mathcal{H}_1$ and $\mathcal{H}_2$ which will now read
\begin{align}
\mathcal{H}_{1,\textrm{int}}=&g_{1\parallel} I_R^z J_{3L}^z+\frac{g_{1\perp}}{2} \left(I_R^+J_3^-+I_R^-J_{3L}^+\right)\nonumber\\
\mathcal{H}_{2,\textrm{int}}=&g_{1\parallel} I_L^z J_{3R}^z+\frac{g_{1\perp}}{2} \left(I_L^+J_{3R}^-+I_L^-J_{3R}^+\right)
\end{align}
with $g_{1\parallel}=g_{1\perp}=g_1$ in the SU(2)-invariant case.
The resulting Hamiltonian is solvable at a particular point, known as a Toulouse point, upon a simple unitary transformation. 
This approach is well known from the theory of the Kondo problem, and was used with great success by 
Emery and Kivelson in the two-channel Kondo problem.\cite{emery-1992} The SU(2) invariant Hamiltonian of 
Eqs.\eqref{SU(2)2+Z2} and \eqref{H1H2} (solved by Andrei {\it et al.} using Bethe ansatz methods\cite{andrei-1998}) and its 
U(1)-invariant version are also related by an irrelevant operator at the Toulouse point. 

A summary of the  Azaria-Lecheminant solution at  the Toulouse point is presented in Appendix \ref{sec:toulouse}.
The upshot of this solution is that the low energy degrees of freedom at the Toulouse point are  a massless boson, 
that we denote by  $\bar \Phi_1$, and  two massless Majorana fermions, denoted by  $\xi^0$ and $\chi^1$ respectively. 
Since the Bose field $\bar \Phi_1$ is at the SU(2)$_1$ radius,  the Luttinger liquid Hamiltonian $\mathcal{H}_{LL}$ is equivalent to a theory of an 
SU(2)$_1$ WZW model with right- and left-moving chiral currents $\bm{\mathcal{J}}_R$ and $\bm{\mathcal{J}}_L$
\begin{align}
 \mathcal{J}_R^\pm&=\frac{1}{2\pi a} e^{\mp i \sqrt{8\pi} \bar \Phi_{1,R}}, & \mathcal{J}^z_R&=\frac{1}{\sqrt{2\pi}} \partial_x \bar \Phi_{1,R}\nonumber\\
  \mathcal{J}_L^\pm&=\frac{1}{2\pi a} e^{\pm i \sqrt{8\pi} \bar \Phi_{1,L}}, & \mathcal{J}^z_L&=\frac{1}{\sqrt{2\pi}} \partial_x \bar \Phi_{1,L}\nonumber\\
  &&&
 \label{effective-SU(2)1}
 \end{align}
In the notation of Appendix \ref{sec:toulouse}, the effective Hamiltonian at the Toulouse point is
\begin{equation}
\mathcal{H}_{TP}=\frac{2\pi u_1}{3} \Big(\bm{\mathcal{J}}_R^2+\bm{\mathcal{J}}_L^2\Big)+ \mathcal{H}_F[\xi^0]+\mathcal{H}_F[\chi^1]
\label{HTP}
\end{equation}
where $\mathcal{H}_F[\xi^0]$ and $\mathcal{H}_F[\chi^1]$ are the Hamiltonians for the free Majorana Fermi fields $\xi^0$ and $\chi^1$, given in Eq.\eqref{HFxi0} 
and Eq. \eqref{HFchi1} of Appendix \ref{sec:toulouse}.
 The effective low energy Hamiltonian of Eq.\eqref{HTP}  is the chiral fixed point of Azaria and Lecheminant. It describes a conformally-invariant system 
 with total central charge $c=2$ (in the spin sector alone), down by one unit with respect to the decoupled system.  
 
 \subsubsection{Gapless Fractionalized Spin Fluid}
 \label{sec:fractionalized-fluid}
 
 To assess the significance of this fixed point to the frustrated Kondo-Heisenberg chain we need to consider the stability of this fixed point. 
 The results of Azaria and Lecheminant will be useful but require some changes given the differences in the two problems. 
 In particular they showed that the corrections to the Toulouse point Hamiltonian that correct for the artificial breaking of the SU(2) symmetry amounts 
 to an irrelevant perturbation.  
 
  We will now see that for $J_1$ weak enough there is a novel gapless fractionalized fluid phase characterized, in the spin sector, 
  by the stable fixed point of Azaria and Lecheminant. 
 In this phase we will find that in the charge sector there is coexistence between the PDW order with the singlet superconductor (and other more conventional orders). 
 However both in the charge and spin sectors the leading operators, which have the smallest scaling dimension and hence the more strongly divergent susceptibilities 
 at low temperatures, are unconventional. In this sense this phase is characterized for having intertwined orders in a fractionalized fluid state.

 As we stressed before in the problem at hand we do not have a coupling between the N\'eel order parameters of the spin chains with each other 
 and with the 1D electronic system. We thus need to only consider the role of the backscattering interactions between the two spin chains 
 (with coupling constant $g_2$ in Eq\eqref{HHKSF}), and the backscattering interactions in each spin chain 
 (with coupling constant $\bar g$ in Eq.\eqref{HHKSF}) and in the 1D electronic system (with coupling constant $\bar g'$ in Eq.\eqref{HHKSF}).  
 We will further assume that the coupling constants $\bar g$ and $\bar g'$ are approximately equal to each other. 
 The results of Ref. [\onlinecite{azaria-2000}] imply that in the low-energy limit 
 both the inter-chain spin-current backscattering interactions and the intra-chain backscattering spin-current interaction (including the 1D electronic system) 
 map onto the operator $\bm{\mathcal J}_R \cdot \bm{\mathcal J}_L$ (although with different effective coupling constants) where $\bm{\mathcal J}_{R,L}$ 
 are the chiral SU(2)$_1$ currents defined in Eq.\eqref{effective-SU(2)1}.
  
Given these considerations we now can write the full Hamiltonian for the charge and spin sectors $\mathcal{H}=\mathcal{H}_c+\mathcal{H}_s$, where $\mathcal{H}_c$ 
is the Hamiltonian for the charge boson $\phi_c$ presented in Eq.\eqref{Hcharge}, and $\mathcal{H}_s$ is the Hamiltonian for the spin sector of Eq.\eqref{HHKSF} 
at the chiral fixed point (the Toulouse point)
\begin{align}
\mathcal{H}=& \frac{v_c}{2K_c} \Pi_c^2+\frac{1}{2}K_c v_c (\partial_x \phi_c)^2\nonumber\\
+&\frac{2\pi u_1}{3} \Big(\bm{\mathcal{J}}_R^2+\bm{\mathcal{J}}_L^2\Big)+ g_{\rm eff} \bm{\mathcal{J}}_R\cdot \bm{\mathcal J}_L\nonumber\\
-& \frac{iv_1}{2}\left( \xi^0_R\partial_x\xi^0_R-\xi^0_L\partial_x\xi^0_L\right)\nonumber\\
-&\frac{iv_1}{2}\left( \chi^1_R\partial_x\chi^1_R-\chi^1_L\partial_x\chi^1_L\right)
\label{HTP+pert+charge}
\end{align}
where $g_{\rm eff}=3 \bar g + 2 g_2$ is the effective coupling constant for backscattering interactions of the SU(2)$_1$ currents of Eq.\eqref{effective-SU(2)1} . 
This interaction  with  coupling constant $g_{\rm eff}$, 
which in abelian bosonization leads to an operator of the form $\cos(\sqrt{8\pi} \bar \Phi_1)$ in the effective Hamiltonian (see Eq.\eqref{effective-SU(2)1}),
is a marginal operator. It is a relevant perturbation for $g_{\rm eff}>0$ and an irrelevant perturbation for 
$g_{\rm eff}<0$. Since $g_2 \sim J_1 >0$ and $\bar g\sim -J_2<0$, the relevance or irrelevance of this interaction depends on the relative strengths of 
these couplings. In the limit that we are considering here $g_{\rm eff}<0$ until the interaction $J_1$ becomes large enough. 

We then conclude that, at least for weak enough interchain interaction $J_1$, the chiral fixed point is perturbatively stable provided $g_{\rm eff}<0$. 
Hence in the regime $J_1\ll J_K \ll J_2$ there is a new stable phase which is characterized by the finite fixed point, which is described by the Toulouse 
Hamiltonian of Eq. \eqref{HTP+pert+charge} (up to irrelevant operators). We will now look at the behavior of the correlators of the frustrated Kondo-Heisenberg chain. 
To this end we need to find the form of the physical observables of the Kondo-Heisenberg chain in this description.

Azaria and Lecheminant gave an identification\cite{azaria-2000} 
at the Toulouse point  for the N\'eel order parameters $\mathbf{N}_1$ and $\mathbf{N}_2$, as well as the SDW order parameter of the 1D electric chain $\mathbf{N}_3$, 
and of the spin currents $\mathbf{J}_1$, $\mathbf{J}_2$ and $\mathbf{J}_3$. 
Here we will adapt these methods to identify at the Toulouse point, in terms of the fields of Eq.\eqref{HTP+pert+charge},
 the observables of the 1D electronic system $\Delta_{SS} (x) $ and $\bm{\Delta}_{TS}(x)$, the singlet and triplet superconductor operators, 
and its CDW and SDW (``N\'eel) order parameters $\rho_{CDW}(x)$ and $\bm{N}_{SDW}(x)$ (both with ordering wave-vectors $2k_F$). We will also need 
the N\'eel order parameters $\bm{N}_1$ and $\bm{N}_2$ of each magnetic chain (both with ordering wave-vector $\pi$), 
and in the composite order parameters $\bm{\Delta}_{TS} \cdot (\bm{N}_1\pm \bm{N}_2)$, which describe a pair-density-wave order parameter $\Delta^{PDW}_\pm$ 
also with wave-vector $\pi$. 
As in the case of the unfrustrated Kondo-Heisenberg chain,\cite{zachar-2001,berg-2010} 
the PDW order parameters involve the degrees of freedom of the 1D electronic system 
and of the antiferromagnetic chains. 

The spin currents $\mathbf{J}_3$ and the SDW (N\'eel) order parameter $\mathbf{N}_3$ of the 1D electronic chain  
at the Toulouse point  are identified (to leading order) with the following operators\cite{azaria-2000}
\begin{align}
J^z_{3,R} &\sim - \mathcal{J}^z_L, &J^\pm_{3,R} &\sim \mathcal{J}^\pm_L 
\nonumber\\
J^z_{3,L} &\sim -\mathcal{J}^z_R, &J^\pm_{3,L} &\sim \mathcal{J}^\pm_R
\nonumber\\
N_3^z &\sim -i \pi a \; \chi^1_R \chi^1_L \mathcal{N}^z,&N_3^\pm &\sim i \pi a\; \chi^1_R \chi^1_L\mathcal{N}^\pm
\nonumber\\
&&&
\label{J3N3-TP}
\end{align}
where the fields $\mathcal{N}^z$ and $\mathcal{N}^\pm$ are given by
\begin{align}
\mathcal{N}^z\sim& -\frac{1}{\pi a} \sin(\sqrt{2\pi} \bar \Phi_1) \nonumber\\
\mathcal{N}^\pm \sim & \frac{1}{\pi a} e^{\pm i \sqrt{2\pi} \bar \Theta_1}\nonumber\\
\mathcal{E} \sim & \frac{1}{\pi a} \cos(\sqrt{2\pi} \bar \Phi_1)
\label{SU(2)1-primaries}
\end{align}
are the components of the (spin-1/2)  primary field $g$ (with scaling dimension 1/2) of the SU(2)$_1$ WZW conformal field theory
\begin{equation}
g \sim \mathcal{E}+i \bm{\mathcal{N}} \cdot \bm{\sigma}
\label{SU(2)1-g}
\end{equation}
where $\bm{\sigma}$ are the three Pauli matrices. 

The spin currents $\mathbf{J}_1$ and $\mathbf{J}_2$ of the spin chains, and the associated N\'eel order parameters $\mathbf{N}_1$ 
and $\mathbf{N}_2$ have the following (leading order) operator identification at the Toulouse point\cite{azaria-2000}
\begin{align}
J_{1R}^\pm \sim & \mathcal{J}^\pm_R  \Big(1+i a \pi   \chi^1_L \xi^0_R\Big)\nonumber\\
J_{2R}^\pm \sim & \mathcal{J}^\pm_R  \Big(1-i a \pi   \chi^1_L \xi^0_R\Big)\nonumber\\
J_{1R}^z=J_{2R}^z=&\mathcal{J}^z_R
\label{J1J2-TP}
\end{align}
where $a$ is the short-distance cuttoff (the lattice spacing). Similar expressions hold for the left moving components (with the $R$ and $L$ labels exchanged). 

The N\'eel order parameters $\mathbf{N}_1$ and $\mathbf{N}_2$
have the following identifications\cite{azaria-2000}
\begin{align}
N_1^z\sim & \mathcal{N}^z (\mu_5\mu_0+\sigma_5\sigma_0)\nonumber\\
N_2^z\sim & \mathcal{N}^z (\mu_5\mu_0-\sigma_5\sigma_0)\nonumber\\
N_1^\pm \sim & \mathcal{N}^\pm (\mu_5\mu_0-\sigma_5\sigma_0)\nonumber\\
N_2^\pm \sim & \mathcal{N}^\pm (\mu_5\mu_0+\sigma_5\sigma_0)
\label{N1N2-TP}
\end{align}
Here $\sigma_0$ and $\sigma_5$, and  $\mu_0$ and $\mu_5$ are respectively  the order and disorder twist field operators associated with the critical Ising 
models with Majorana fields $\xi^0$  and $\chi^1$ (see Appendix \ref{sec:toulouse}).

We will now turn to the identification at the Toulouse point of the order parameters involving the charge sector of the frustrated Kondo-Heisenberg chain. 
They are the  CDW order parameter $\rho_{CDW}$ of the 1D electronic system,
\begin{equation}
\rho_{CDW}\sim \frac{1}{\pi a} \chi^1_R \chi^1_L \; e^{-i \sqrt{2\pi} \phi_c}\; \cos(\sqrt{2\pi} \bar \Phi_1)
\label{rho-cdw}
\end{equation}
 the singlet and triplet superconductor order  parameters $\Delta_{SS}$ and $\bm{\Delta}_{TS}$ of the 1D electronic system,
 \begin{align}
 \Delta_{SS} \sim & \frac{1}{\pi a} \chi^1_R \chi^1_L \; e^{-i \sqrt{2\pi}  \theta_c}\; \cos(\sqrt{2\pi} \bar \Phi_1)
 \label{singletSC}\\
\bm{ \Delta}_{TS} \sim & \frac{1}{\pi a} \chi^1_R \chi^1_L \; e^{-i \sqrt{2\pi} \theta_c}\; \bm{\mathcal{N}}
 \label{tripletSC}
 \end{align}
 and the PDW order parameters the the frustrated Kondo-Heisenberg chain $\Delta^{PDW}_\pm=\bm{\Delta} \cdot (\mathbf{N}_1\pm \mathbf{N}_2)$,
 \begin{align}
\Delta^{PDW}_+ \sim \sigma_5 \sigma_0 \; e^{-i\sqrt{2\pi}\theta_c}
\label{DeltaPDW+}\\
\Delta^{PDW}_- \sim \mu_5 \mu_0 \; e^{-i\sqrt{2\pi}\theta_c}
\label{DeltaPDW-}
\end{align}
where we used that the Majorana mass term $i\chi^1_R \chi^1_L \sim \varepsilon_1$ where $\varepsilon_1$ is the energy operator of the Ising model 
(the scaling dimension $1$ relevant ``thermal'' operator at the critical point) and the operator-product expansion (OPE) $\sigma_5 \; \varepsilon_1 \sim \sigma_5$ 
(likewise the disorder operator satisfies $\mu_5 \varepsilon_1 \sim \mu_5$). Similarly the Majorana bilinear $i \xi^0_R \xi^0_L$ is identified with the energy density 
operator $\varepsilon^0$ of another critical Ising model and satisfies the same OPEs with its order and disorder operators, $\sigma_0$ and $\mu_0$.

Finally, the charge-$4e$ uniform superconducting order parameter $\Delta_{4e}$ (already discussed in Section \ref{sec:J2llJ1}, Eq.\eqref{Delta4e}) 
which is the square of the 
PDW order parameters (given in Eq.\eqref{DeltaPDW+} and Eq.\eqref{DeltaPDW-}), is in turn identified with $\Delta_{4e}\sim \exp(i 2 \sqrt{2\pi} \theta_c)$. This operator has 
scaling dimension $2K_c$. Notice that here we used the fact that to leading order the Ising twist fields $\sigma$ and $\mu$ to leading order fuse into the identity operator. 
It is easy to see that the next-to-leading term involves the energy density operator $\varepsilon$ of the Ising model. Since the energy density operator 
$\varepsilon$ has scaling dimension $1$, 
the leading correction  to the
PDW order parameter has scaling dimension $4K_c+1$. 

We can now characterize the phase controlled by the non-trivial (finite) chiral fixed point represented by the effective Hamiltonian of 
Eq.\eqref{HTP+pert+charge} in the regime $g_{\rm eff}<0$. At this fixed point (and in the low energy regime of the entire phase which it controls) 
the four fields $\phi_c$, $\bar \Phi_1$, $\xi^0$ and $\chi^1$ are massless and hence critical. The only effect of the marginally irrelevant perturbation 
$\bm{\mathcal{J}}_R\cdot \bm{\mathcal{J}}_L$ is to induce logarithmic corrections to the correlators (except those that involve conserved currents). 

\paragraph{Order parameters of the spin sector}: 
The scaling dimensions of the spin currents $\mathbf{J}_{k,R,L}$ (with $k=1,2,3$) is $1$. The scaling dimension of the SDW order parameter $\mathbf{N}_3$ 
(given in Eq.\eqref{J3N3-TP})  is $\frac{3}{2}$ since this operator is the product of the Ising energy density operator $\varepsilon_1$ (which has scaling dimension $1$) 
and of the SU(2)$_1$ primary field $\bm{\mathcal{N}}$ (which has scaling dimension $\frac{1}{2}$). 
Instead, the N\'eel order parameters $\mathbf{N}_{1}$ and $\mathbf{N}_{2}$ 
(given in Eq.\eqref{N1N2-TP}   of the spin chains is $3/4$ since each disorder (and order) operator has dimension $\frac{1}{8}$. 
In contrast, at the decoupled fixed point of Eq.\eqref{HHKSF} with $g_1=0$, the scaling dimensions of all three N\'eel order parameters is $\frac{1}{2}$. 
Thus, the Kondo exchange coupling has caused the scaling dimensions of the three N\'eel order parameters to increase considerably and the critical exponents  
of their correlation functions are now $\eta_1=\eta_2=3/2$ and $\eta_3=\eta_{SDW}=3$.
 
We should also consider  the spin singlet composite operators $\mathfrak{N}_\pm=\mathbf{N} \cdot (\mathbf{N}_1\pm \mathbf{N}_2)$. 
Both operators have ordering wave vectors $Q_\pm=\pi-2k_F$. Since these operators are spin singlets, they 
can be interpreted as  composite CDW order parameters. 
A similar operator also exists in the weakly frustrated case.\cite{berg-2010}
In addition, the operator $\mathfrak{N}_-$ is odd under the exchange of the two magnetic chains (as is also the odd PDW order parameter $\Delta^{PDW}_- $).
It is easy to see that at the Toulouse point  the operators $\mathfrak{N}_\pm$ are identified 
(to leading order) with the operators $\mathfrak{N}_+\sim \mu_5\mu_0$ and $\mathfrak{N}_- \sim \sigma_5 \sigma_0$, respectively. 
Hence, both operators $\mathfrak{N}_\pm$ have scaling dimension 
$\frac{1}{4}$ and hence their critical exponent is $\eta_{\mathfrak{N}_\pm}=\frac{1}{2}$. 
Therefore in this phase, the dominant order in the spin sector is given by the composite order parameter 
$\mathfrak{N}_\pm$, followed by the N\'eel orders of the magnetic chains and by the SDW order parameter of the 1D electronic system.

\paragraph{Order parameters of the charge sector}: 
We can also read-off that the  scaling dimensions of operators that involve the charge sector. The scaling dimensions of the CDW order parameter is 
$\frac{1}{2}\left(3+\frac{1}{K_c}\right)$ at the Toulouse point, while the scaling dimension of the singlet superconducting order parameter  is $\frac{1}{2}(K_c+3)$ 
(as it is for the triplet superconductor), and  the scaling dimensions of the PDW orders is $\frac{1}{2}\left(K_c+\frac{1}{2}\right)$ and for the charge-$4e$ 
superconductor is $2K_c$. Their critical exponents are, respectively, given by
$\eta_{SS}=\eta_{TS}=K_c+3$, $\eta_{CDW}=3+\frac{1}{K_c}$, $\eta_{PDW}=K_c+\frac{1}{2}$ and $\eta_{4e}=4K_c$.

From this analysis we conclude that in this phase all of the orders are present, with the composite CDW order parameters $\mathfrak{N}_\pm$ 
and the PDW superconducting order parameter $\Delta_{PDW}$ having the correlation function that decay more slowly with distance, and hence 
are the dominant orders in this phase. In this sense this phase is analogous to the PDW phase of the weakly frustrated regime. 
However, in this phase these unconventional orders also coexist with the more conventional N\'eel orders of the chains and in the electronic system, 
as well as the singlet (and triplet) superconducting orders, the CDW order and the charge $4e$ superconducting order. 
Therefore this phase has a gapless charge sector (with central charge $c=1$) and a partially gapped spin sector (with total central charge $c=2$).
This fractionalized phase has a large number of orders that are comparably strong and hence are {\em intertwined} (instead of {\em competing} with each other). 
This is a feature that is shared with the PDW phase of the weakly frustrated regime. In contrast, in the dimerized phase the frustrated Heisenberg chain 
and the 1D electronic system are simply decoupled at low energies, and the allowed orders are more conventional.

\subsubsection{Quantum phase transition to the PDW phase}
\label{sec:QPT-PDW}

We will now consider the case in which $g_{\rm eff}>0$ in which case the operator $\bm{\mathcal{J}}_R \cdot \bm{\mathcal{J}}_L$ becomes marginally relevant. 
This is a Kosterlitz-Thouless type transition. In this phase the field $\bar \Phi_1$ becomes pinned at the values $\left(n+\frac{1}{2}\right) \sqrt{\frac{\pi}{2}}$, 
and its fluctuations are massive. Consequently there is a gap in most (but not all) the spin degrees of freedom. This leads to a further reduction of the central charge of the spin sector of at least from $c=2$ to $c=1$, while the charge sector remains massless and has central charge $c=1$. We will see however that the spin sector is fully gapped and that the resulting phase is equivalent to the PDW phase we discussed in the weak frustration regime.

In the regime in which $g_{\rm eff}>0$, operators such as $\cos(\sqrt{2\pi}\bar \Phi_1)$ have vanishing expectation values in this phase, while operator 
such as $\sin(\sqrt{2\pi} \bar \Phi_1)$ do not.  Moreover, in this phase the dual field $\bar \Theta_1$ fluctuates wildly and vertex operators of the dual field have 
vanishing expectation values. Therefore in this phase the field $\bar \Phi_1$ is massive and decouples from the low-energy physics. 
The effective Hamiltonian for this phase has the same form as the effective Hamiltonian at the Toulouse point,  Eq.\eqref{HTP}, except that the SU(2)$_1$ 
currents $\bm{\mathcal{J}}_{R,L}$ have been projected out. 
Hence, in this phase the remaining massless degrees of freedom are the charge boson $\phi_c$ and the Majorana fermions $\xi^0$ 
and $\chi^1$. We will see, however, that these Majorana fermions actually become massive.

These
considerations allow us to determine the behavior of the observables of interest. 
 It is easy to see that all the order parameters that are not expressed as composite operators have exponentially decaying correlators 
and are not condensed. This includes, the singlet and triplet superconducting order  parameters $\Delta_{SS}$ and $\bm{\Delta}_{TS}$, 
the SDW and CDW order parameters $\mathbf{N}_3$ and  $\rho_{CDW}$ of the 1D electronic system, 
and the N\'eel order parameters $\mathbf{N}_1$ and $\mathbf{N}_2$ of the spin chains 
(see Eqs. \eqref{J3N3-TP}, \eqref{N1N2-TP}, \eqref{rho-cdw}, \eqref{singletSC} and \eqref{tripletSC}). 
All these operators have exponentially decaying correlation functions. Nevertheless several composite operators have power-law correlation functions.

\paragraph{Order parameters of the spin sector}: The only operators from the spin sector that have power-law correlations in this phase are the spin-singlet operators 
$\mathfrak{N}_\pm$. As we saw, these operators are given by products of Ising twist operators. Hence also in this phase these operators have scaling dimension $1/4$ 
and critical exponent $\eta_{\mathfrak{N}_\pm}=1/2$, and their correlation functions oscillate with wave vector $\pi-2k_F$.

\paragraph{Order parameters of the charge sector}: The only operators in this phase with power-law correlations in the charge sector are the PDW operators of 
Eqs. \eqref{DeltaPDW+} and \eqref{DeltaPDW-}, and the charge $4e$ uniform superconductor. Since, as we will see below, the Majorana fermions $\xi^0$ and $\chi^1$ become massive, in this phase the scaling dimensions of $\Delta^{PDW}_\pm$ and $\Delta_{4e}$ are the same as in the PDW phase.

What is the difference between this regime and the PDW phase? So far looks almost identical to the PDW phase of the weakly frustrated Kondo-Heisenberg chain, albeit with different critical exponents if the Majorana fermions $\xi^0$ and $\chi^1$ remain massless. 
Microscopically, the main difference between this phase and the PDW phase of the weakly frustrated regime is that the latter phase in the symmetry of 
exchanging the two spin chains is broken, whereas superficially here it seems to be unbroken. 

Let us examine this question more closely.  From the structure of the effective Hamiltonian,  and of the observables, we see that what distinguishes this possible phase from the PDW phase is whether
 the product of Ising twist fields $\sigma_0 \sigma_5$ has (or has not) a non-vanishing expectation value.  
This is achieved by combining both Majorana fermions into a single Dirac fermion $\psi=\frac{1}{\sqrt{2}} (\xi^0+i \chi^1)$ 
and generating a Dirac mass term or, equivalently, a Majorana mass for both Majorana fermions, i.e. the energy density operators of the two Ising models. 
In particular, in the phase in which $\langle \sigma_0\sigma_5 \rangle \neq 0$, the disorder operators have vanishing expectation value, 
$\langle \mu_0 \mu_5\rangle=0$, and vice versa (by Kramers-Wannier duality). Consequently across the phase transition the symmetric PDW operator, $\Delta_+^{PDW}$ has short-range correlations, whereas the operator $\Delta_-^{PDW}$ becomes the PDW order parameter of the weakly frustrated case. 

However, for the expectation value of the product of twist fields to be different from zero the 
 the two remaining Majorana fermions $\xi^0$ and $\chi^1$ must become massive. Such a mass term cannot be generated in the fractionalized fluid phase. 
 However, a term of this form is generated from nominally irrelevant operators (whose scaling dimension at the Toulouse fixed point is $3$ and larger) 
 once the operator $\bm{\mathcal{J}}_R \cdot \bm{\mathcal{J}}_L$ becomes marginally relevant and the field 
 $\bar \Phi_1$ becomes massive. 
 That this is correct can be seen form the identification of the spin currents, {\it e.g.} Eq.\eqref{J1J2-TP}.  
 These operators exist since at the microscopic level the symmetry of exchanging the two magnetic chains is explicitly broken. 
 However, the operators that break this symmetry are irrelevant at the Toulouse point and hence do not destabilize the gapless fractionalized spin fluid phase. Irrelevant operators that break the symmetry of a system at a fixed point are known as dangerous irrelevant operators. Such operators do not change the universality class of the phase transition but change the nature of the resulting stable phase.
Moreover, the backscattering current interactions of the spin currents of the two magnetic chains and of the 1D electronic system, Eq.\eqref{HHKSF},  
contain operators that break this symmetry but which are irrelevant at the Toulouse point and at the quantum phase transition out of the fractionalized 
gapless spin fluid phase. Thus, the resulting phase is a PDW phase due to the existence of these dangerous irrelevant operators which are irrelevant at criticality 
but which lower the symmetry of the stable phase.

\section{Phase Diagram and Conclusions}
\label{sec:conclusions}

\begin{figure}[hbt]
\begin{center}
\includegraphics[width=0.4\textwidth]{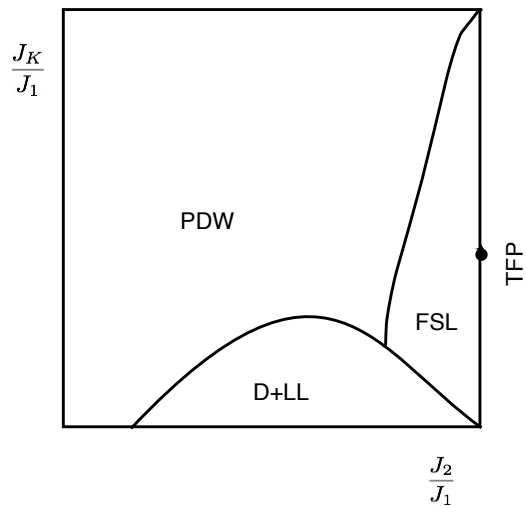}
\end{center}
\caption{Schematic phase diagram for the frustrated Kondo-Heisenberg chain. Here $J_1$ and $J_2$ respectively represent the nearest-neighbor and next-nearest neighbor exchange interactions and  $J_K$ the Kondo exchange interaction between the the spins of the frustrated Heisenberg chain and the 1D electronic system. PDW denotes the pair-density-wave superconducting phase, FSL the fractionalized spin liquid (controlled by the Toulouse fixed point (TFP)), and D+LL the phase with a dimerized Heisenberg chain coexisting with a Luttinger liquid in the 1D electronic system. The quantum phase transition between the D+LL phase and the PDW and FSL phases are most likely first order (see text). The phase boundary between the FSL and the PDW phases is of the Kosterlitz-Thouless class, for $J_K$ weak enough, but may be first order for $J_K$ large enough. We have not found any evidence of a phase transition at $J_K\to \infty$.  }
\label{fig:diagram}
\end{figure}

In this paper we discussed the structure of the phase diagram of the frustrated Kondo-Heisenberg chain. In addition to the PDW phase discussed by 
Berg, Fradkin and Kivelson,\cite{berg-2010} we find that quantum frustration of the Heisenberg chain leads to the complex phase diagram shown in Fig. \ref{fig:diagram}.
%%%%%%%%%%%%%%%%%%%%%
%%%%%%%%%%%%%%%%%%%%%%
In the preceding sections we found that the phase diagram has the following phases: 
\begin{enumerate}
\item
A PDW phase in the weak frustration regime, $J_1\gg J_2$. This phase is controlled by a stable fixed point at large $J_K$ and weak $J_2$. 
\item
 For $J_2> J_{2,c}$ and $J_K$ small there is a dimerized phase of the spin chain (with a spin gap)  coexisting with a spin-1/2 Luttinger liquid in the 1D electronic system. The  phase boundary between the dimerized spin chain+Luttinger liquid (D+LL) is most likely weakly first order since the system at $J_K=0$ and $J_2=J_{2,c}$ is at a fixed point with two marginally relevant operators.
 \item
 A fractionalized spin liquid (FSL) phase with a gapless (fractionalized) spin fluid coexisting with a decoupled charge sector of the 1D electronic system. We investigated the nature of the quantum phase transition between the dimerized and the PDW phase and concluded that it most likely a first order transition. We also investigated the stability of the gapless spin fluid phase (characterized by a stable fixed point of the Toulouse form at finite $J_K$) and the nature of its correlation functions. We also showed that, for weak enough $J_K$,  there is a Kosterlitz-Thouless (KT) type quantum phase transition from the gapless fractionalized spin fluid to another phase which, due to a dangerous irrelevant operator, it is physically equivalent to the PDW phase of the weakly frustrated regime. This KT transition most likely becomes a first order transition at large enough $J_K$ since in for $J_K \to \infty$ the system is unstable to a flow either towards the Toulouse fixed point (TFP) of towards the stable fixed point of the PDW phase.
We have not found evidence a direct phase transition between the dimerized phase and the gapless fractionalized spin fluid. This suggests that either the PDW phase ``sneaks in'' between this phase and the dimerized phase even in the strong frustration regime or, what is far more likely, that there is a direct first order phase transition.
\end{enumerate}

One of our motivations for looking at the role of frustration in the spin chain was to find out if a magnetic mechanism can give rise to an incommensurate PDW phase in a system without attractive interactions. As we noted in the introduction such a mechanism appears to be forbidden in 1D systems with full SU(2) spin rotational invariance. Nevertheless such incommensurate phases are possible if the SU(2) spin symmetry is explicitly broken down to its U(1) subgroup by a magnetic anisotropy interaction in the spin chain.

\begin{acknowledgments}
A. Dobry thanks the Institute for Condensed Matter Theory of the University of Illinois for hospitality and E. Fradkin thanks the Department of Physics of the 
Faculty of Exact and Natural Sciences of the University of Buenos Aires for hospitality. 
This work was supported in part by the National Science Foundation, under grant DMR-1064319 (EF) at the University of Illinois,
and by the U.S. Department of Energy, Division of Materials Sciences under Award No. 
DE-FG02-07ER46453 through the Frederick
Seitz Materials Research Laboratory of the University of Illinois (AJ), by Programa Ra{\'\i}ces of the Ministry of Science and Technology or Argentina (EF), by PIP  11220090100392 of CONICET (AD), and  PICT R 1776 of the ANPCyT (AD). A.J. thanks the University of Oklahoma and Prof. B. Uchoa for their kind support.
\end{acknowledgments}

\appendix
\section{Solution at the Toulouse Point}
\label{sec:toulouse}

The Hamiltonian of Eqs. \eqref{SU(2)2+Z2} and \eqref{H1H2} can be  treated using abelian bosonization as follows. 
We will first describe the spin degrees of freedom of the electronic system in terms of a Bose field that we will denote by $\varphi$ and whose dual field is $\vartheta$. 
Their chiral components are, as usual, $\varphi_R=(\varphi+\vartheta)/2$ and $\varphi_L=(\varphi-\vartheta)/2$ respectively. 
The chiral fields $\varphi_{R,L}$ are defined at the $SU(2)_1$ compactification radius, such that the chiral spin currents of the electronic system, 
$J^\pm_{3,s}$ and $J_{3,s}$ are given by the expressions of the form of Eq.\eqref{SU(2)1} in terms of the chiral fields $\varphi_s$ (with $s=R,L$),
\begin{equation}
J_{3,s}^\pm=\frac{1}{2\pi a} e^{\mp i \sqrt{8\pi} \varphi_s}, \quad J_{3,s}^z=\frac{1}{\sqrt{2\pi}} \partial_x \varphi_s
\label{SU(2)1-1DEG}
\end{equation}

We will also need an expression for the chiral SU(2)$_2$ currents $I^a_{s}$ of Eq.\eqref{SU2_2} in abelian bosonization. This is accomplished by combining two of the Majorana Fermi 
fields, $\xi_{1,s}$ and $\xi_{2,s}$ into a single free Dirac (complex) Fermi field $\psi_s=(\xi_{s}^2+i\xi_{s}^1)/\sqrt{2}$. 
The  free Dirac fermion has a representation in terms of a chiral Bose fields $\phi_R$ and $\phi_L$
\begin{align}
\psi_R=\frac{1}{\sqrt{2}} \left(\xi_R^2+i\xi_R^1\right)&=\frac{1}{\sqrt{2\pi a}} e^{i \sqrt{4\pi} \phi_R}\nonumber\\
\psi_L=\frac{1}{\sqrt{2}} \left(\xi_L^2+i\xi_L^1\right)&=\frac{1}{\sqrt{2\pi a}} e^{-i\sqrt{4\pi} \phi_L}
\label{fermi-bose}
\end{align}
In this language the chiral SU(2)$_2$ currents of Eq.\eqref{SU2_2} are given by
\begin{align}
	I^\pm_R &=\frac{i}{\sqrt{\pi a}} \xi^3_R  F_R e^{\pm i \sqrt{4\pi}\phi_R}, \qquad I^z_R = \frac{1}{\sqrt{\pi }} \partial_x\phi_R
	\nonumber\\
	I^\pm_L &= \frac{i}{\sqrt{\pi a}} \xi^3_L F_L e^{\mp i \sqrt{4\pi}\phi_L}, \qquad I^z_L= \frac{1}{\sqrt{\pi }} \partial_x \phi_L	
	\label{SU2_2-bosonized}
\end{align}
where $F_{R,L}$ are two (anticommuting) Klein operators. It is easy to see that the SU(2)$_2$ chiral currents have scaling dimension $1$ (as they should) 
and obey an SU(2)$_2$ Kac-Moody current algebra.

In this formulation the Hamiltonians $\mathcal{H}_1$ and $\mathcal{H}_2$ of Eq.\eqref{H1H2} take the form
\begin{align}
	\mathcal{H}_1 =& v_1 (\partial_x\phi_R)^2 + v_2 (\partial_x\varphi_{L})^2 + \frac{g_{1\parallel}}{\sqrt{2}\pi} \partial_x\phi_R\partial_x\varphi_{L}  \nonumber\\
		 -&\frac{i v_1}{2} \xi^3_R\partial_x\xi^3_R  + \frac{i g_{1\perp}}{2\pi\sqrt{\pi a}} \xi^3_{R} {F}_R \cos(\sqrt{4\pi}\phi_R+\sqrt{8\pi}\varphi_{L}) 
		\label{H1}\\
	\mathcal{H}_2 =& v_1 (\partial_x\phi_L)^2 + v_2 (\partial_x\varphi_{R})^2 + \frac{g_{1\parallel}}{\sqrt{2}\pi} \partial_x\phi_L\partial_x\varphi_{R}\nonumber\\
	+&\frac{i v_1}{2} \xi^3_L\partial_x\xi^3_L + \frac{i g_{1\perp}}{2\pi \sqrt{\pi a}} \xi^3_L {F}_L \cos(\sqrt{4\pi}\phi_L +\sqrt{8\pi}\varphi_{R}) 
	\label{H2}
\end{align}
The derivative couplings in the Hamiltonian $\mathcal{H}_1$ in Eq.\eqref{H1} can be eliminated by means of a canonical transformation of the form
\begin{equation}
\begin{pmatrix}
\phi_R
\\
\varphi_L
\end{pmatrix}
=\begin{pmatrix}
\cosh \alpha & -\sinh \alpha\\
-\sinh \alpha & \cosh \alpha
\end{pmatrix}
\begin{pmatrix}
\bar \Phi_{1R}\\
\bar \Phi_{2L}
\end{pmatrix}
\label{canonical-H1}
\end{equation}
provided we choose 
\begin{equation}
\tanh (2\alpha)=\frac{g_{1\parallel}}{\pi \sqrt{2} (v_1+v_2)}
\end{equation}
Similarly, the derivative couplings in the Hamiltonian $\mathcal{H}_2$ of Eq.\eqref{H2} can be eliminated by a canonical transformation of the same form 
(with the same value of $\alpha$)
\begin{equation}
\begin{pmatrix}
\varphi_R
\\
\phi_L
\end{pmatrix}
=\begin{pmatrix}
\cosh \alpha & -\sinh \alpha\\
-\sinh \alpha & \cosh \alpha
\end{pmatrix}
\begin{pmatrix}
\bar \Phi_{2R}\\
\bar \Phi_{1L}
\end{pmatrix}
\label{canonical-H2}
\end{equation}

For general values of $g_{1\parallel}$ the cosine terms in Eqs.\eqref{H1} and \eqref{H2} have non-trivial scaling dimension. 
The Toulouse point is defined as the value of $g_{1\parallel}$ for which the scaling dimension of the cosine operators is $1/2$. This happens for 
\begin{equation}
\tanh \alpha^{TP}=\frac{1}{\sqrt{2}}, \qquad g_{1\parallel}^{TP}=\frac{4\pi}{3}(v_1+v_2)
\end{equation}
At the Toulouse point the Hamiltonian $\mathcal{H}_1+\mathcal{H}_2$ of Eq.\eqref{H1H2} becomes
\begin{align}
	\mathcal{H}_1 +\mathcal{H}_2=	& u_1 \left[ (\partial_x\bar \Phi_{1R})^2 + (\partial_x\bar \Phi_{1L})^2 \right]\nonumber\\
	                                                      +& u_2 \left[ (\partial_x \bar \Phi_{2R})^2 +(\partial_x\bar \Phi_{2L})^2\right] \nonumber\\
	                                                       -&\frac{i v_1}{2} \left( \xi^3_R\partial_x\xi^3_R - \xi^3_L\partial_x\xi^3_L\right) \nonumber\\
	                                                       +&\frac{i g_\perp}{2\pi \sqrt{\pi a}} \Big[ \xi^3_{R} {F}_R \cos(\sqrt{4\pi}\bar \Phi_{2L})\nonumber\\
	                                                       &\;\;\;\;\;\;\; +\xi^3_L {F}_L \cos(\sqrt{4\pi}\bar \Phi_{2R}) \Big]
	                                                       \label{H1+H2}
\end{align}
where $u_1=(2v_1-v_2)/3$ and $u_2=(2v_2-v_1)/3$ are the renormalized velocities at the Toulouse point. 

Since the cosine operators of Eq.\eqref{H1+H2} have scaling dimension $1/2$, they can be refermionized (exactly as in the case of the two-channel 
Kondo problem\cite{emery-1992}).  Let $\chi(x)$ denote a Dirac Fermi field, whose chiral components are 
\begin{equation}
\chi_R=\frac{1}{\sqrt{2}} (\chi_{2,R}+i \chi_{1,R}), \quad \chi_L=\frac{1}{\sqrt{2}} (\chi_{2,L}+i\chi_{1,L})
\end{equation}
 respectively, where $\chi_{j,s}$ are the two Majorana components ($j=1,2$) with both chiralities ($s=R,L$) of the Dirac fermion. 
In turn, the chiral Dirac fermions $\chi_{R}$ and $\chi_{L}$ are related to the chiral bosons $\bar \Phi_{2,R}$ and $\bar \Phi_{2,L}$ by an expression of the same form 
as in Eq.\eqref{fermi-bose}, i.e.
\begin{equation}
\chi_R=\frac{1}{\sqrt{2\pi a}} \bar F_R e^{i\sqrt{4\pi} \bar \Phi_{2,R}}, \qquad \chi_L=\frac{1}{\sqrt{2\pi a}} \bar F_L e^{-i \sqrt{4\pi} \bar \Phi_{2,L}}
\end{equation}
where $\bar F_R$ and $\bar F_L$ are two Klein factors.

 In this basis, at the Toulouse point the Hamiltonian $\mathcal{H}_s$ of Eq.\eqref{SU(2)2+Z2} can be written as a sum of four decoupled Hamiltonians respectively 
 consisting of  the Luttinger liquid Hamiltonian for the Bose field $\bar \Phi_1$ (and its dual field $\bar \Theta_1$), the Hamiltonians for the free Majorana fermions $\xi^0$ 
 and  $\chi^1$, and the Hamiltonian for the Majorana fermions $\xi^3$ and $ \chi_2$ coupled to each other through a mass term,
\begin{equation}
\mathcal{H}_s=\mathcal{H}_{LL}[\bar \Phi_1]+ \mathcal{H}_F[\xi^0]+\mathcal{H}_F[\chi^1]+\mathcal{H}_F[\xi^3,\chi^2]
\label{Hs-split}
\end{equation}
where %$\mathcal{H}_{LL}[\bar \Phi_1]$, $\mathcal{H}_F[\xi^0]$, $\mathcal{H}_F[\chi^1]$, and $\mathcal{H}_F[\xi^3,\chi^2]$ are given by
\begin{align}
\mathcal{H}_{LL}[\bar \Phi_1]=& \frac{u_1}{2}\left[(\partial_x\bar \Phi_1)^2 + (\partial_x\bar \Theta_1)^2 \right]
\label{HLL-barPhi1}
\\
	\mathcal{H}_F[\xi^0]=& -\frac{iv_1}{2}\left( \xi^0_R\partial_x\xi^0_R-\xi^0_L\partial_x\xi^0_L\right)
	\label{HFxi0}
	\\
	\mathcal{H}_F[\chi^1]=& -\frac{iv_1}{2}\left( \chi^1_R\partial_x\chi^1_R-\chi^1_L\partial_x\chi^1_L\right)
	\label{HFchi1}
	\\
\mathcal{H}_F[\xi^3,\chi^2]=& -\frac{i v_1}{2} \Big( \xi^3_R\partial_x\xi^3_R - \xi^3_L\partial_x\xi^3_L\Big) 
	\nonumber\\
				     &-\frac{i u_2}{2} \left( \chi^2_R\partial_x\chi^2_R - \chi^2_L\partial_x\chi^2_L\right)
				     \nonumber\\
				     & + i m \left( \xi^3_{R} \chi^2_L -  \chi^2_R \xi^3_L \right)
				     \label{HFxi3+chi2}
\end{align}
where we have defined the mass $m=g_{1\perp}/(2\pi a)$.
The Hamiltonian $\mathcal{H}_F[\xi^3,\chi^2]$ of Eq.\eqref{HFxi3+chi2} represents a theory of two free Majorana fermions coupled through the mass term shown 
in the last term on the right hand side of Eq.\eqref{HFxi3+chi2}. The spectrum of single-particle states of  the Hamiltonian $\mathcal{H}_F[\xi^3,\chi^2]$ is
\begin{equation}
	E_{\pm\pm}(k) = \pm u_1 k \pm \sqrt{\left(\frac{v_1+v_2}{2}\right)^2 k^2 + m^2}
\end{equation}
This spectrum  is  massive with a single-particle gap of $m\sqrt{1-\left(u_1/v_+\right)^2}<m$, where $u_1=(2v_1-v_2)/3$ and $v_+=(v_1+v_2)/3$.

Therefore, at the Toulouse point the Majorana fermions $\xi^3$ and $\chi^2$ become massive and decouple from the other low energy degrees of freedom, 
represented by the massless boson  $\bar \Phi_1$, and the massless Majorana fermions  $\xi^0$ and $\chi^1$.

The 2D classical Ising model and, equivalently, the 1D (quantum) Ising model in a transverse field,
is equivalent to a theory of Majorana fermions with a Majorana mass term which is tuned to zero at criticality.\cite{Schultz-1964,Zuber-1977}
In the language of 2D conformal field theory, the Ising order parameter which we will denote by $\sigma$,  and its Kramers-Wannier dual, the disorder operator 
which we will denote by $\mu$, respectively have non-vanishing expectation values on each of the order and disorder 
phases of the Ising model.\cite{Kadanoff-1971,fradkin-1978}  In the fermionic language the two phases correspond to the two
possible signs of the Majorana mass term (also known as the energy operator of the Ising model, which we will denote by $\varepsilon$).\cite{difrancesco-1997}
In addition of labeling the order and disorder phases, the order and disorder operators twist the boundary conditions of the Majorana fermion from periodic 
to anti-periodic, 
and hence are called twist-fields.\cite{Dixon-1987} The scaling dimensions of the order and disorder operators at the Ising critical point, where the 
Majorana fermion is massless, is $1/8$. In the construction outlined in this Appendix we encountered six Majorana fermions, 
$\xi^0$, $\xi^1$, $\xi^2$, $\xi^3$, $\chi^2$ and $\chi^1$. Therefore we will also have six associated twist fields $\sigma^0, \ldots \sigma^5$ (in this order), 
and their associated Kramers-Wannier duals $\mu^0, \ldots,\mu^5$. These order and disorder twist fields enter in the definition of the observables  of the 
frustrated Kondo-Heisenberg chain in this description.

\end{document}